\documentclass[showpacs,showkeys,11pt,
preprint,preprintnumbers,nofootinbib,
groupedaddress,superscriptaddress,amsmath,amssymb]{revtex4}
\usepackage{amsfonts}
\usepackage{amsmath}
\usepackage{graphics}
\usepackage[export]{adjustbox}
\usepackage{epsfig}
\usepackage{url}
\usepackage{multirow}
\usepackage{feynmp}

\newcommand {\be}{\begin{equation}}
\newcommand {\ee}{\end{equation}}
\newcommand {\ba}{\begin{eqnarray}}
\newcommand {\ea}{\end{eqnarray}}

\begin{document}
\title{Double and Triple Higgs Boson Production at Future Linear Colliders}

\pacs{12.60.Fr, 
      14.80.Fd  
}
\keywords{Charged Higgs, MSSM, LHC}
\author{Ijaz Ahmed}
\email{Ijaz.ahmed@cern.ch}
\affiliation{Riphah International University, Sector I-14, Hajj Complex, Islamabad Pakistan}
\author{Taimoor Khurshid}
\email{taimoor.khurshid@iiu.edu.pk}
\affiliation{International Islamic University, H-10, Islamabad}


\begin{abstract}
In the present work such processes which act as a source of double Higgs production as well as triple Higgs production at the future linear colliders are analyzed within the general Two Higgs Doublet Model and Minimal Supersymmetric Model at different center of mass energies. The main objective is to compute the cross-sections of Higgs pair production process $H^{+}H^{-}$ as well as triple Higgs boson processes like $H^{+}H^{-}H^{0}$, $H^{+}H^{-}h^{0}$ at $\sqrt{s}$ = 1.5 TeV and $\sqrt{s}$ = 3 TeV. The cross-sections of both types of processes are also compared with the general Two Higgs Doublet Model as well as minimal Super Symmetric Model. It is observed that for double charged Higgs production process (H${}^{+}$H${}^{-}$) the resulting cross-section values are almost same in both the models, while in the case of triple Higgs production process, several orders of enhancement in cross-sections is observed in 2HDM as compared to MSSM. In addition, neutral Higgs pair production $A^{0} H^{0}$ and $A^{0} h^{0}$ and triple neutral Higgs production $h^{0}h^{0}A^{0}$ is studied in 2HDM type-I at the future linear ${\mu}^{+}$${\mu}^{-\ }$ collider by scanning energies from $\sqrt{s}$ = 0.5 TeV to $\sqrt{s}$ = 3 TeV. It is observed that the cross-section for neutral Higgs pair production $A^{0} h^{0}$ process is comparatively greater than that of neutral Higgs pair production $A^{0} H^{0}$ process. Similarly for neutral Higgs Triple production $h^{0}h^{0}A^{0}$, an enhancement is seen in cross-sections with increase of tan${\beta}$.  
\end{abstract}
\maketitle
\section{Introduction}
The SM \cite{lab1,lab2,lab3,lab4} is most significant considerable and conspicuous Quantum Field Theory (QFT) dealing with the fundamental particles of nature and also explains the mechanism in which they interact with each other. It explains successfully all the known particles in existence to a remarkable degree of accuracy. That's why the SM is also regarded as the model of almost everything in the universe, as it explains successful experimental results. Standard model assumes a single complex Higgs Boson ($H^{0}$) \cite{lab5,lab6,lab7} also called as God particle. Higgs boson is responsible to generate the mass of all fermions \cite{lab8} and gauge bosons \cite{lab9} through Yukawa coupling terms \cite{lab10}. Moreover the God particle, Higgs boson is the only experimentally discovered scalar boson, having mass 125.09$GeV/c^{2}$, charge and spin zero. It has no intrinsic spin, that's why it is classified as a boson. Higgs boson was discovered by the ATLAS and CMS collaborations \cite{lab11,lab12,lab13} in 2012 from collisions between protons (protons were accelerated at 99.99\% of the speed of light) at the world's largest, most powerful and highest-energy particle accelerator machine, Large Hadron Collider (LHC) \cite{lab14,lab15,lab16}. Currently, the SM is the best description of the micro-world, as it predicted many particles before their discovery. Moreover, the standard model is the most authentic and frequently used model, but yet there are many mysterious that SM does not explain, such as, it does not describe the particle's quantum numbers, no predictions about mass spectrum, \cite{lab17} it also ignores gravity, no information about neutrinos mass, does not describe the generations of fermions and gives no information about the extended Higgs sector (Charged Higgs $H^{+}$,$H^{-}$) \cite{lab18}. We know that SM contains only a scalar Higgs having spin zero responsible for generating masses of all other particles. But it does not explain the multiple Higgs. So study of charged Higgs is not possible through the SM.
These unsolved questions in the standard model motivate physicists to continue searches and experiments in order to discover new physics that lie ahead of the SM at higher TeV scale. The incoming decade is expected to be an exciting period of physics beyond the SM (High Energy Physics). The main objective of physics beyond the SM is to resolve the problems within the SM as well as to investigate and search the properties of the basic constituents of the matter with each other and with those of the fundamental interactions in nature in a range of energy never reached, at several tera-electron volts (TeV) energies \cite{lab19}. A number of possible models investigated by both the detectors ATLAS and CMS, beyond the SM such as two-Higgs doublet model \cite{lab20,lab21,lab22} and Minimal super-symmetric standard model \cite{lab24,lab25,lab26} include various extensions of the Standard Model. Both these models are extensively used by the physicists and are most favorable for the study of the charged Higgs bosons \cite{lab18} as well as for neutral Higgs bosons \cite{lab27} because these models are dealing with multiple Higgs bosons. Basically the purpose of this paper is to search the sources of charged Higgs pair production ($H^{+}H^{-}$), triple Higgs production ($H^{+}H^{-}H^{0}$), ($H^{+}H^{-}h^{0}$) neutral Higgs pair production($A^{0} H^{0}$) as well as ($A^{0} h{0}$) and neutral Higgs Triple production ($h^{0} h^{0} A^{0}$) at  future linear colliders under 2HDM and its type-II (MSSM) for various values of center of mass energies.

\section{The Two Higgs Doublet Model}
The 2HDM \cite{lab20,lab21,lab22} is the most widely-studied model in which a second complex doublet is added to the scalar sector. The 2HDM contains multiple charged Higgs bosons as in CP-conserving case there are two ($H^{+} H^{-}$) charged Higgs scalar bosons, a CP-even electrically neutral Higgs $H^{0}$, and also CP-odd electrically neutral or (Pseudo Scalar) Higgs $A^{0}$. The 2HDM is the most frequently used model and studied theoretically as well as limited experimentally. This model is the simplest beyond the SM extension of the Higgs mechanism in the electroweak symmetry breaking \cite{lab28} which makes originate naturally in the scalar sector of different theories. It is a minimal extension of SM Higgs sector, having extra scalar doublet that includes many physical neutral as well as charged Higgs field. 2HDM contains $\Phi_{1}$ and $\Phi_{2}$ two Higgs doublet instead of one.
\begin{equation}
\phi_i = \begin{pmatrix} \phi_i^{+}\\ \phi_i^{-} \end{pmatrix} 
\label{eq1}
\end{equation}
Hence there are total eight degrees of freedom that are used to generate the mass of intermediate vector bosons. In some other cases, after symmetry breaking, three Goldstone bosons give longitudinal modes of the $W^{\pm }$ and $Z$ bosons, that become massive and there will remain only five Higgs bosons, among these three light, heavy and pseudoscalar $h^{0}$, $H^{0}$, $0^{are}$ are neutral and two $H^{+}$, $H^{-}$ are charged scalar Higgs bosons.
The most general renormalizable and gauge invariant 2HDM potential V($\Phi_{1}$, $\Phi_{2}$) \cite{lab28,lab29} is defined and expressed in the 8-dimensional space of the Higgs field. As it is a combination that is important of electroweak invariant combinations such as $\Big( \phi_{1}^{\dagger} \phi_{1} \Big) $, $\Big( \phi_{2}^{\dagger} \phi_{2} \Big) $, $\Big( \phi_{1}^{\dagger} \phi_{2} \Big) $, $\Big( \phi_{2}^{\dagger} \phi_{1} \Big) $ and $\Big( \phi_{i}^{\dagger} \phi_{j} \Big) $ where i $\wedge j = {\rm 1}\wedge {\rm 2}$. The most general 2HDM potential equation is 
\begin{center}
\begin{equation}
\begin{aligned}
V = m_{11}^2 \phi_{1}^{\dagger} \phi_{1} + m_{22}^2 \phi_{2}^{\dagger} \phi_{2} - m_{12}^2 \phi_{1}^{\dagger} \phi_{2} + H.c. + \frac{1}{2} \lambda_1 (\phi_{1}^{\dagger} \phi_{1})^2 + \\ \frac{1}{2} \lambda_2 (\phi_{2}^{\dagger} \phi_{2})^2 +   \lambda_3 (\phi_{1}^{\dagger} \phi_{1} ) \times (\phi_{2}^{\dagger} \phi_{2} ) +  \lambda_4 (\phi_{1}^{\dagger} \phi_{2} ) \times (\phi_{2}^{\dagger} \phi_{1} ) + ( \frac{1}{2} \lambda_5 (\phi_{1}^{\dagger} \phi_{2} )^2 + \\ \lambda_6 (\phi_{1}^{\dagger} \phi_{1} ) \times (\phi_{1}^{\dagger} \phi_{2} ) +  \lambda_7(\phi_{2}^{\dagger} \phi_{2} ) \times (\phi_{1}^{\dagger} \phi_{2} ) + H.c. )
 \label{eq2}
 \end{aligned}
 \end{equation}
\end{center}
The most general scalar potential includes 14 real free parameters \cite{lab30}. The $m_{11}^2$, $m_{22}^2$, $\lambda_1$, $\lambda_2$, $\lambda_3 \wedge \lambda_4$ are real parameters and the complex parameters are $m_{12}^2$, $\lambda_5$, $\lambda_6 \wedge \lambda_7$. In 2HDM this potential plays vital role for the symmetry breaking\cite{lab27} and stability, moreover after the electroweak symmetry breaking \cite{lab27} it is responsible with the interaction terms from the kinetic terms in order to generate the masses of gauge bosons. However, these large numbers of free parameters make the characterization of the symmetry breaking for the regions in parameters space very complicated. The potential is not unique in contrast with the SM, as each set of parameters lead to different interactions, mass, and Feynman rules etc. Therefore 2HDM is governed by the choice of Higgs potential parameters and moreover by the Yukawa couplings \cite{lab33} of the two scalars Higgs doublets to generations of fermions \cite{lab8}. \\
To avoid flavor changing neutral currents (FCNC) both the doublets $\Phi_{1}$ and $\Phi_{2}$ should have different quantum numbers with each other. The easiest way is to impose the $Z_{2}$ symmetry \cite{lab32}. As the $Z_{2}$ is a discrete symmetry \cite{lab32} which is often imposed in order to avoid the tree level FCNCs. Basically FCNCs is a type action of some weaker forces and in this phenomena flavour of particle is changed but electric charge remains unchanged. As discrete $Z{2}$ symmetry lead to $\lambda _{{\rm 6}}$ ,$\lambda _{7} = 0$ by imposing $Z_{2}$ symmetry extra terms can be reduced in potential equation.\\ 
There are various versions of 2HDM with different choices of the parameters that have interesting characteristics. As 2HDM can introduce flavor changing neutral currents and there are four types \cite{lab33} of Yukawa interactions are possible depending on which type of fermions couples to which doublets $\Phi$.
 \begin{table}[ht]
  \begin{center}
  \begin{tabular}{|c|c|c|c|}
  \hline
  Model &  &  &  \\ \hline
  Type-I & $\Phi$${}_{2}$ & $\Phi$${}_{2}$ & $\Phi$${}_{2}$ \\ \hline
  Type-II & $\Phi$${}_{2}$ & $\Phi$${}_{1}$ & $\Phi$${}_{1}$ \\ \hline 
  Lepton-specific & $\Phi$${}_{2}$ & $\Phi$${}_{2}$ & $\Phi$${}_{1}$ \\ \hline
  Flipped & $\Phi$${}_{2}$ & $\Phi$${}_{1}$ & $\Phi$${}_{2}$ \\ \hline 
  \end{tabular}
  \caption{ Particular Types of 2HDM}
  \label{tab1}
  \end{center}
  \end{table}
In type-I, there is only one Higgs doublet generate the mass of all fermions and intermediate vector boson, \cite{lab24,lab25} as the second doublet $\Phi_{2}$ contributes via mixing (only $\Phi_{2}$ couples to fermions). So all the fermions only couples to one doublet. Hence the Higgs phenomenology depicts similar behavior to SM. The 2HDM-II is like Minimal Super-symmetric standard Model \cite{lab23,lab24,lab25} likes model as only a single doublet couple to up type quarks and another one to down type quarks and lepton. In type-II doublet one $\Phi$${}_{1}$ couples  to down type, doublet $\Phi$${}_{2}$ couples to fermions of up type. Natural flavor conservation is featured, and its phenomenology is just like the 2HDM-I. However, the SM couplings are not only shared through mixing but also can be shared through the Yukawa structure. In 2HDM type-III, due to the presence of flavor changing neutral interactions that's why it is different from others models, and required a special type of suppression mechanism e.g. Yukawa coupling \cite{lab33} imposed dedicated texture. Moreover $\Phi$${}_{1}$ couples to down quarks and $\Phi$${}_{2}$ couples to up quarks and also down leptons. In 2HDM type-IV, up type frictionally charged particle quarks couples to one doublet and down type quarks couples to other doublet.
\section{The Minimal Super-Symmetric Standard Model}
The Minimal super-symmetric standard model (MSSM) \cite{lab23,lab24,lab25} is the simplest super symmetric extension of the standard model, with R-parity conservation and soft super symmetry (SUSY) \cite{lab34,lab35} breaking proposed in 1981. It is a particular type-II in 2HDM that realizes the super symmetry. It only considers the minimum amount of new interactions and new particle states accordant with phenomenology. Basically MSSM realizes important phenomena of SUSY and only consider minimum number of some new particle states. The SUSY pairs the elementary particles fermions with the spin 1 particle bosons. In this way every particle of SM contains its corresponding particle which is known as super-partner. Moreover, it makes no assumption about the soft super-symmetry breaking mechanism and introduces the minimum number of new particles as the super-symmetry pair's bosons with fermions, so every particle of SM has a super-partner yet not discovered. A kind of space-time symmetry SUSY is a possible candidate for undiscovered particle physics, however, seen as an elegant solution to many current problems in particle physics, if confirmed accurate which could resolve various field where existing theories are believed to be broken. A super symmetrical extension to the SM, the MSSM would overcome as well as solve the major problems of particle physics. The MSSM is still under investigation by the LHC experiments \cite{lab14,lab15,lab16}.
The MSSM Higgs sector is slightly complex than that of SM. However it is analogous to 2HDM \cite{lab20,lab21}  as the SM has one Higgs doublet only which gives masses to the fermions and intermediate vector bosons, but the MSSM requires to Higgs doublets $\Phi_{1}$ and $\Phi_{2\ }$in order to produce the mass of both the down as well as up type fractionally charged quarks. In addition, we also need two Higgs doublets to ensure anomaly cancellation. As these two Higgs doublets $\Phi_{1}$ and $\Phi_{2}$ having eight degrees of freedom between them, as three are used to give the mass for the $W^{\pm}$ and $Z$ bosons \cite{lab9}. Hence five degrees of freedom and five physical Higgs bosons left behind. The MSSM Higgs sector is CP-conserving at the lowest order and contains $h$, $H$, $A$, $H^{+}$, $H^{-}$. The light Higgs h and heavy Higgs H are CP-even, with $M_{h} \mathrm{<} M_{H}$, the A Higgs boson (pseudoscalar Higgs) is CP-odd, however, $H^{\pm}$ are charged. 
At tree level \cite{lab36}  two Higgs doublets are required in super symmetric theories (as in Minimal super-symmetric standard model) in order to produce the masses for the charged leptons and up type as well as down type quarks.
$$\begin{pmatrix} H_{u}^{+} \\ H_{u}^{0} \end{pmatrix} \qquad \qquad \qquad  \qquad  \qquad \begin{pmatrix} H_{d}^{+} \\ H_{d}^{0} \end{pmatrix}$$ 
Couples to up-fermions			$\qquad$		Couples to down-fermions.\\
Hence two doublets are needed to give all particles mass. In particular, there is an important relationship that relates the charged Higgs boson mass with the CP-odd Higgs boson mass $A$(pseudoscalar boson), and mass of $W$ intermediate vector boson. So, in this case, the tree-level Higgs mass is given.
\begin{equation}
\label{eq3}
m_{H^\pm}^2 = m_{A}^2 + m_{W}^2
\end{equation}

\section{Future Linear Particle Colliders}
The future linear colliders \cite{lab37}  are the basically purposed concept of the future particle accelerators. The physicists believed that future linear colliders of particles will help to resolve the current problems in particle physics, concerning the fundamental theories that governs interactions and basic forces between the elementary particles, nature of dark matter, also the structure of time and space and relation of quantum mechanics and theory of relativity where the existing theories of physics and knowledge about them are not sufficient to understand and not applicable till now. Moreover, they will search for new theories of physics which are still considered as incomplete and also all purposed phenomena's at higher energies, such as at several TeV levels. Two main future linear colliders are $e{+}$ $e^{-}$ (ILC and CLIC) \cite{lab38,lab39}  and ${\mu}^{+}$ ${\mu}^{-}$. These two types of colliders will generate precise results and will operate with various collision energies. The most important benefit of these future colliders is that it is easily possible to stage experiment at some other energy throughout the lifetime experiment.
\subsection{Future Linear $e{+}$ $e^{-}$ colliders}
Actually, both ILC as well as CLIC are $e^{+}$ $e^{-}$ future purposed linear colliders \cite{lab40,lab41,lab42}, In order to explore some new frontiers of energy. ILC is under planning stage to have 500 GeV collision energy initially and the possibility for later on promoting to 1 TeV or 1000 GeV. It would collide the beams of positrons with electrons at almost speed of light in order to give precise experimental results. As the early proposed place for this colliders were CERN (Europe) \cite{lab43,lab44} Fermi-lab (USA) and Japan. However the government of Japan is willing to contribute half of the total cost, so Japan is considered as the most suitable location for its instalment. 
It is about 31 km long having high luminosity \cite{lab45} based on about 1.3 GHz advanced accelerating super conducting radio frequency technology. Similarly, CLIC is also $e^{+}$ $e^{-}$ future linear collider that plans to work at collision energy up to 3 TeV. It will work at intermediate energy stages so such energies are to be measured by ongoing experimental work at the world largest machine LHC \cite{lab14,lab15,lab16}. It would have to beam acceleration technique and the staged construction will produce collision energy up to 3 TeV. It would collide positrons beams with electron and currently, it is considered as an option for multi TeV future collider. It is purposed to built at CERN and it will be between 11 km to 50 km long.
\subsection{ Future Linear ${\mu}^{+}$ ${\mu}^{-}$ collider}
 A future linear ${\mu}^{+}$ ${\mu}^{-}$ collider \cite{lab13,lab14,lab15} is basically proposed technology of particle accelerator. It will collide beams of muon ${\mu}^{-}$ with beams of anti-muon ${\mu}^{+}$ for its operation to observe experimental results collision energy of 4 TeV. Actually, the muon is leptons which are produced during energetic collisions of cosmic rays and also can be created in particle accelerators. It is proposed to be constructing at Fermi-lab. The MAP is basically Muon Accelerator program, \cite{lab46}  started in 2010 to promote the idea of the muon collider. The main aim of MAP is to explore the new theories and undiscovered phenomena of physics which are under investigation by the physicists. This program plays an important role in the development of linear muon anti-muon collider.  
 
\subsection{Advantages of ${\mu}^{+}$ ${\mu}^{-}$ collider over $e{+}$ $e^{-}$ collider} 
The main convenient advantages of muons, rather than electrons, for a lepton collider are: 
\begin{enumerate}
 \item The synchrotron radiation, \cite{lab47}  (such radiations emitted by charged particles when they are circulated in the presence of magnetic field) that powers the electron colliders of high to be linear, is ( Eq.\ref{eq4}) that powers high energy electron colliders to be linear, is (see Eq.\ref{eq4} gives the energy lost in the form of synchrotron radiation)
  \begin{equation}
   \label{eq4}
   \Delta V_{turn} \propto \frac{E^{4}}{RM^4} \propto \frac{E^{3} B}{M^4}
  \end{equation}
Inversely proportional to mass the fourth power it is very negligible in the case of muon colliders. Hence a muon collider can be circular. Practically speaking this implies it can belittler. So there is most important advantage of muon colliders to use muon beams, because the rate at which the synchrotron radiations emit is inversely related with m2. Hence conclusion is that greater is the mass of particle (leptons) used in collider smaller are the factor of synchrotron radiations.
 \item The luminosity \cite{lab46} of a $\mu$ collider is shown by a similar equation (Eq. \ref{eq5}) 
  \begin{equation}
    \label{eq5}
	L = \frac{1}{4 \pi E} \frac{N}{\sigma_x} \frac{P_{beam}}{\sigma_y} n_{collisions}
  \end{equation}
as given above for an electron-positron collider, however, there are two significant changes: a) The classical radius $r_o$ is presently that for the $mu$ and is multiple times smaller. b) the total collisions a bunch which can make $n_{collisions}$ is never again 1, yet is currently constrained just by the $\mu$ lifetime and winds up identified with the normal twisting field in the muon collider ring having  
  \begin{equation} 
     \label{eq6}
    n_{collisions} \approx 150 \quad B_{ave} 
  \end{equation} 
The average field of 6 Tesla $n_{collisions}$ $\approx$ 900. These two factors give muons an on a basic luminosity preferred of higher than 105. As a conclusion, for the equivalent luminosity, the required beam power, spot sizes, remittances, and energy spread are far less in $\mu^{+}\mu^{-}$ colliders than in $e^{+}e^{-}$ instruments of the comparable energy.
 \item  The suppression of synchrotron radiation actuated by the contrary group (beamstrahlung) \cite{lab48} permits the utilization of beams with lower momentum and energy spread, and QED radiation is diminished.
  \item  As muon-anti muon colliders would be best in terms of luminosity and power consumption. So it will generate more accurate and precise results than electron-positron colliders.
  \item  It is considered most favorable collider for calculating the cross-sections of leptons, because the factor of energy spread is negligible. Moreover muon-anti muon collider is such an ideal and most significant technology which is proposed to extend the energy frontier about several TeV range. However electron positron linear collider will allow physicists to search new particles by producing collisions up to 1 TeV. But in case of  muon-anti muon colliders physicists would be able to produce collisions up to 3-4TeV. In this way they can get the answers of some unsolved question and understand phenomena's of physics which are not known yet.
 \item Muon anti-muon collider also has an advantage to couples strongly by s-resonance with Higgs mechanism. That's why these proposed linear colliders are considered best for measuring the cross-section. As s-channel Higgs generation \cite{lab49}  is upgraded by a factor of $(m_{\mu} / m_{e})^{2} \approx 40000$. This joined with the small momentum spreads would permit increasingly exact determination of Higgs masses, as well as widths and branching ratios. However, there are issues with the utilization of muons:
 \item  Muons can be produced from the pions decay made higher energy particle such as protons impinging on a target. Moreover, so as to acquire enough muons, a high power proton source is needed with extremely effcient capture of the pions, and muons from their decay.
 \item  The choice of completely polarized muons is conflicting with the need for efficient collection. Polarizations just up to 50 \% are determined, and some loss of luminosity is inescapable ($e^{+}$ $e^{-}$ machines can spellbind the $e^{-}$s up to $\approx$ 85 \%).
\end{enumerate}

\subsubsection{ Double Charged Higgs Production at Future Linear Colliders}

The production of double charged Higgs Bosons has been largely studied in the minimal super symmetric SM \cite{lab24,lab24,lab25,lab26} in the context of future linear $e^{+}$ $e^{-}$ \cite{lab40,lab41,lab42} and $\mu^{+}\mu^{-}$ \cite{lab13,lab14,lab15} colliders. Such kinds of analysis mostly two types of charged Higgs decay are adopted. The first one is $H^{\pm}$ $\to$ $\tau \nu$ and the other one is $H^{\pm}$ $\to$$\tau b$. However, this channel has weak point as it is circumscribed by the collider's CM energy and also a charged Higgs having the mass greater than $\sqrt{s} $/2 cannot be generated unless a negligible amount due to the production of off-shell is considered. The figure 1 represents all the possible Feynman diagrams at the leptons colliders for the double Higgs production.
\begin{figure}[h]
\centering
\includegraphics[width=8cm]{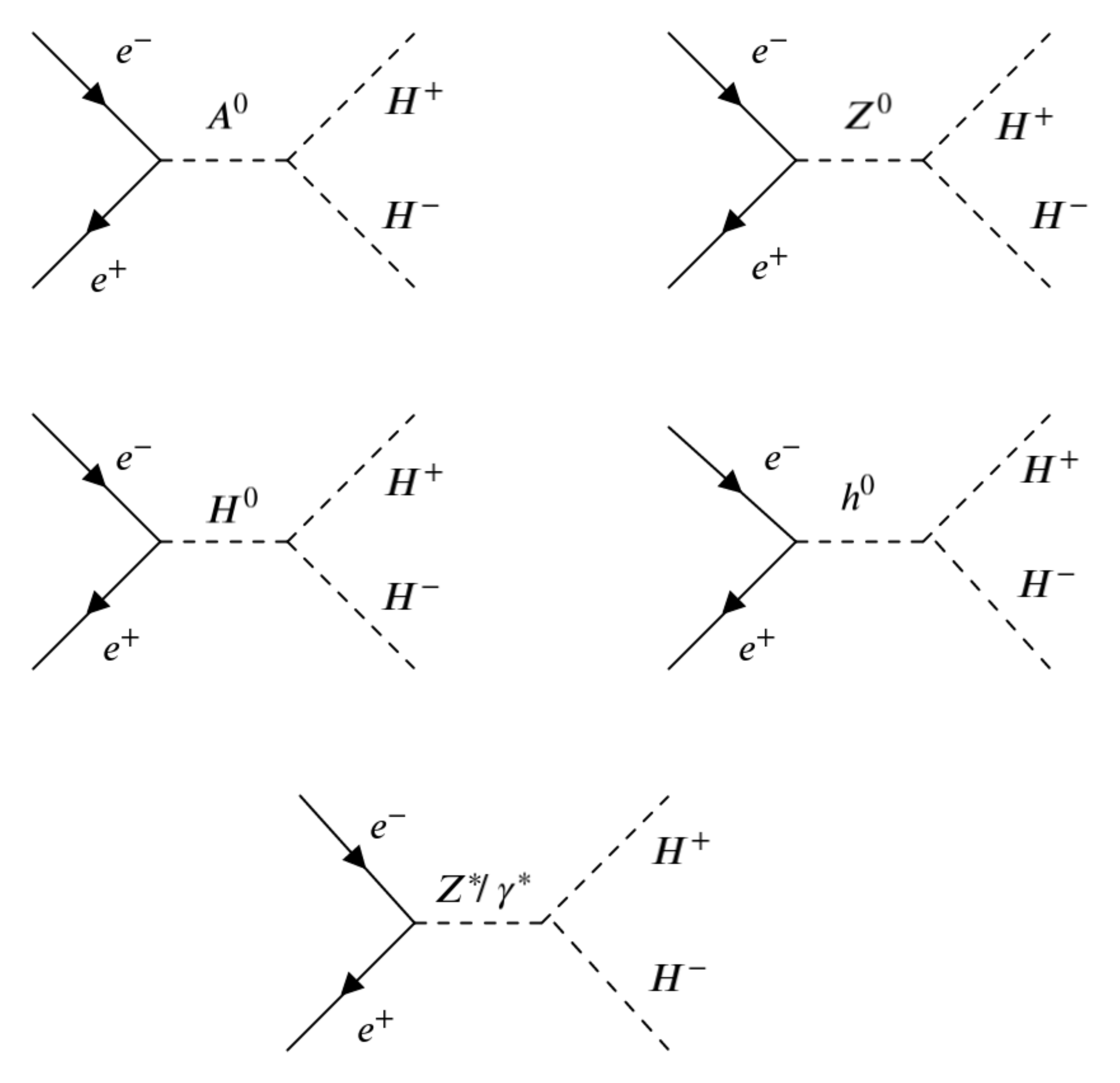}
\caption{ The possible Feynman Diagrams for double Higgs production at lepton collider }
\label{fig1}
\end{figure}

As the table \ref{tab2} depicts the total value of cross-section in fb for the double Higgs production at (1.5 TeV and 3 TeV) CM energy. These processes are analyzed under two different models, the first one is 2HDM and the second one is MSSM under leptons collider (muon anti muon collider). Similarly all double Higgs production processes are studied for the various values of charged Higgs mass at muon anti-muon colliders. The values of cross-section for pair wise Higgs production are listed in table \ref{tab2} at various charged Higgs mass and CM energies. 

\begin{center}
  \begin{table}[ht]
     \centering	
\begin{tabular}{|c|c|c|c|c|c|}  \hline 
Process & $m_{H} \pm$ GeV & $\sigma_{max} (1.5TeV)$ MSSM & $\sigma_{max}$ (3TeV) MSSM & $\sigma_{max\ }$(1.5TeV) 2HDM & $\sigma_{max}$(3TeV) 2HDM \\  \hline 
$\mu^{+}$ $\mu^{-}$$\mathrm{\to}$H${}^{+}$ H${}^{-}$ & 500 & 5.2733 & 2.7401 & 5.2735 & 2.7420 \\ \hline 
$\mu^{+}$ $\mu^{-}$$\mathrm{\to}$H${}^{+}$ H${}^{-}$ & 600 & 2.7036 & 2.5173 & 2.7038 & 2.5184 \\ \hline 
$\mu^{+}$ $\mu^{-}$$\mathrm{\to}$H${}^{+}$ H${}^{-}$ & 700 & 0.5316 & 2.2620 & 0.5324 & 2.2628 \\ \hline 
${\mu}^{+}$ $\mu^{-}$$\mathrm{\to}$H${}^{+}$ H${}^{-}$ & 800 & NP & 1.9687 & NP & 1.9698 \\ \hline 
$\mu^{+}$ $\mu^{-}$$\mathrm{\to}$H${}^{+}$ H${}^{-}$ & 900 & NP & 1.6640 & NP & 1.6653 \\ \hline 
$\mu^{+}$ $\mu^{-}$$\mathrm{\to}$H${}^{+}$ H${}^{-}$ & 1000 & NP & 1.3444 & NP & 1.3458 \\  \hline 
  \end{tabular}
  \caption{  Total cross-section in (fb) of double Higgs at $\sqrt{s} $ =1.5TeV \& $\sqrt{s} $=3TeV Higgs at $\sqrt{s} $ =1.5TeV \& $\sqrt{s}$=3TeV}
  \label{tab2}
  \end{table}
\end{center}

These cross-section values for double Higgs production which are calculated at muon anti muon collider are inversely related with the charged Higgs mass. As from the above calculations when the mass of charged Higgs is increased, in the result cross-section values decreased. Here NP is abbreviated as not possible due to small energy.

\begin{figure}[h]
    \centering
\includegraphics[width=8cm]{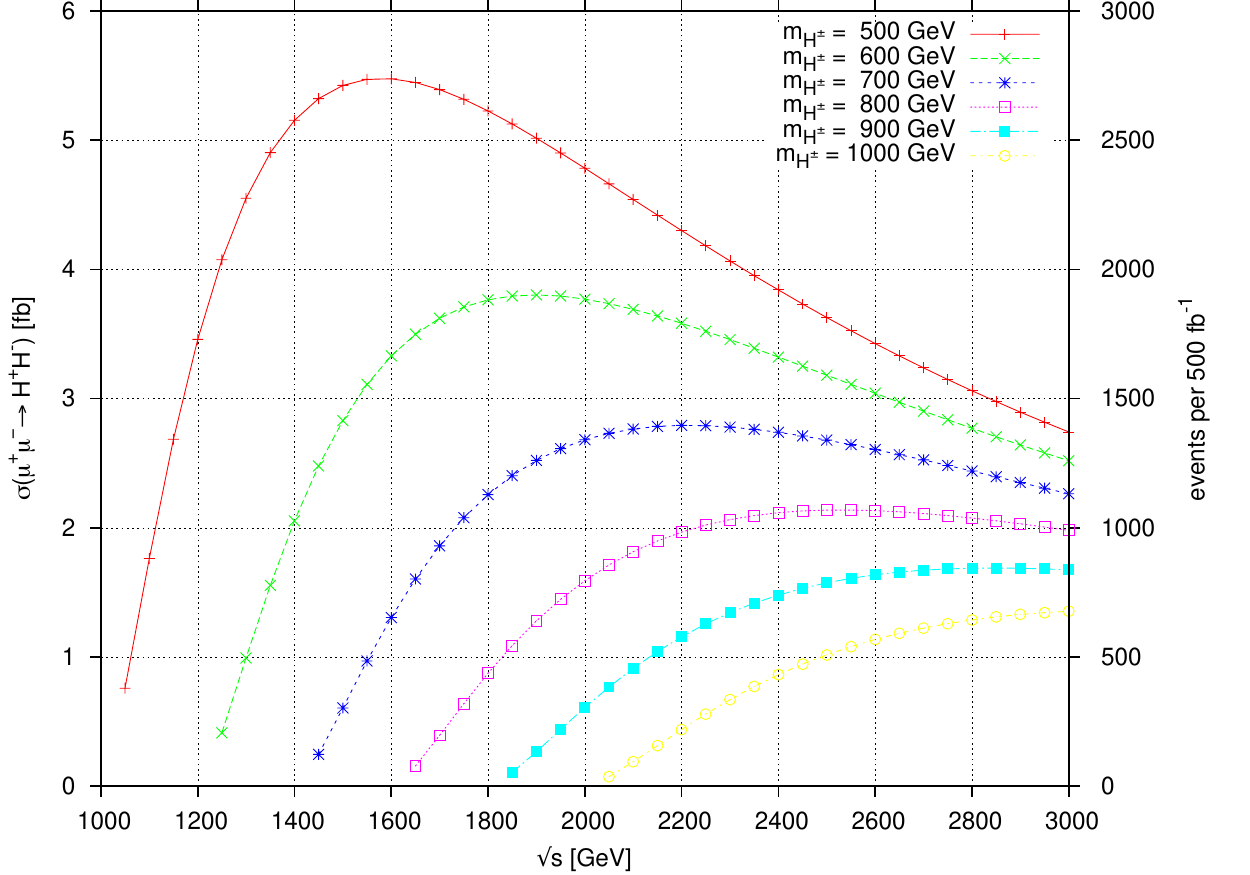}
\caption{ Double Charged Higgs ($H^{+} H^{-}$) pair production cross-section in (fb) at Liner Muon Colliders under 2HDM or MSSM.}
\label{fig2}
\end{figure}

The figure \ref{fig2} shows the total cross section and for ($\mu^{+} \mu^{-} \to H^{+} H^{-}$) which is plotted (as a function of s in GeV) in order to generate the Higgs boson masses in type II of 2HDM \cite{lab20,lab21}. As this cross-section of double charge Higgs production does not depend on the values of $m_{A}$ and $\tan{\beta}$. The figure \ref{fig2} shows that the corresponding rates of production attain few fb achieving thousands of events per fb${}^{-1}$. This graphs also shows that the total cross-section values in fb are function of CM energy, for the double Higgs production at the tree-level under Two Higgs Doublets Model or Minimal Super-symmetric Standard Model. These processes are considered at the muon-anti muon collider. As the cross-section values in fb are represented in y-axis and different values of center of mass energy in GeV are listed in x-axis. Similarly the values of luminosities are also given parallel to that of cross-sections values. Different curves represent the values of cross-section at different values of center of mass energies for various mass of charged Higgs. As when the mass of charged Higgs is about 500GeV then a curve is obtained indicating higher values of cross-section at various values of center of mass energy. On the other hand when the charges Higgs mass is increased then curves are obtained but showing lower values of cross-section e.g. when mass of charged Higgs is about 1000GeV then we get a curve indicating the lowest values of cross-section at various values of center of mass energies. This means that cross-section values are inversely related with mass of charged Higgs and also cross-section is a function of CM energies. Another observation is that number of events per 1000fb${}^{-1}$(luminosities) values also decreases for higher mass of the charged Higgs \cite{lab18}.
As the Feynman diagrams represent an improvement in the rate of production. In the figure \eqref{__4_2_} the Higgs bosons diagrams show the couplings, as well as the contributions of s-channel neutral Higgs Bosons, leptons and the t-channel, charged Higgs Bosons.

\subsubsection{Cross-Section Comparison in MSSM and 2HDM for Double Charged Higgs Production}

The MSSM as well as the 2HDM are basically most favorable model beyond the SM. Both are extended form of SM. The MSSM is actually special type-II of 2HDM model \cite{lab20,lab21}. Both of these models are extensively used by the physicists for the study of charged Higgs Bosons. As these models have multiple Higgs Bosons and act as sources of charged Higgs Bosons \cite{lab18}. That is why physicists preferred these models for study of charged Higgs \cite{lab18}. The total cross-section values in both the models are listed in table 3.

\begin{center}
  \begin{table*}[ht]
     \centering	
\begin{tabular}{|c|c|c|c|c|c|c|}  \hline 
\textbf{Type of Collider} & \textbf{Z}/$\boldsymbol{\gamma}$ & \textbf{MSSM} $\boldsymbol{h^{0}}$/$\boldsymbol{H^{0}}$ & \textbf{Total} & \textbf{Z/$\boldsymbol{\gamma}$} & \textbf{2HDM} \newline $\boldsymbol{h^{0}}$/$\boldsymbol{H^{0}}$ & \textbf{Total} \\  \hline 
$\mu^{+}$$\mu^{-}$ & 2.73 & 1.40x10${}^{-6}$ & 2.73 & 2.74 & 1.41x10${}^{-6}$ & 2.74 \\  \hline 
e${}^{+}$ e${}^{-}$ & 2.73 & 1.27x10${}^{-13}$ & 2.73 & 2.74 & 7.61x10${}^{-16}$ & 2.74 \\  \hline 
  \end{tabular}
  \caption{ \label{tab3} Total cross-section in (fb) of double charged Higgs at $\sqrt{s} $ = 3TeV under MSSM \& 2HDM.}
  \end{table*}
\end{center}

 As table\ref{tab3}  shows the cross-section values are almost equal in MSSM as well as in 2HDM for double Higgs production at specific value of CM energy.  The reason is that coupling of charged Higgs in pair wise charged Higgs \cite{lab18} is almost similar in both of these models. Hence it is very difficult to differentiate between 2HDM and its special type-II the MSSM, because these models same cross-section values at specific CM energy and charged Higgs mass. So the Z/$\gamma$ mediating diagrams are dominating. In both the models .e. 2HDM \cite{lab20,lab21} and MSSM \cite{lab22,lab23,lab24}the resulting cross section almost remains the same. The table\ref{tab3} depicts the contributions of various diagrams in the overall production cross section on specific points in the parameter space. 

\section{ Neutral Higgs Pair Production}
The neutral Higgs boson pair production (2H where 2H = A${}^{0}$H${}^{0}$, A${}^{0}$h${}^{0}$) has been analyzed in large detail of type-I 2HDM. Such type of processes cannot be proceeding in the stranded model because it only deals with a single scalar Higgs (H${}^{0}$). So we can say that in future linear colliders, if we detect sizable rate in the final state of (2H) then it would be an apparent sign of physics BSM. In the production of neutral Higgs pairs production, two types of processes are considered for study in the framework of 2HDM type-I. These two processes of neutral Higgs pair productions are analyzed in the context of most significant future linear colliders at various CM energies from 500 GeV to 3000 GeV. Both the processes of neutral Higgs pair productions $\mu^{+}$ $\mu^{-\ }$$\mathrm{\to}$ A${}^{0}$H${}^{0}$ as well as $\mu^{+}$ $\mu^{-}$ $\to$ $A^{0} h^{0}$ are studied at proposed linear muon anti-muon ( $\mu^{+}$ $\mu^{-}$  )colliders. The main purpose of this analysis is to understand the neutral Higgs pair production process and also the sources of neutral Higgs in 2HDM type-I (type-I is basically SM like scenario) this theoretical study is done by assuming a number of benchmark points in 2HDM \cite{lab20,lab21}  type-I parameter space. Moreover the evidence of predicted pseudo-scalar or CP-odd Higgs (A) and heavy Higgs (H) as well as light or SM Higgs (h) is investigated. The cross-section values in fb are computed at specific values of various parameters. \\
The figure 3 shows the possible Feynman diagrams of neutral Higgs pair production for $\mu^{+}$ $\mu^{-}$$\mathrm{\to}$ A${}^{0}$H${}^{0}$ as well as $\mu^{+}$ $\mu^{-}$ $\mathrm{\to}$ A${}^{0}$h${}^{0}$ processes at the future linear $\mu^{+}$ $\mu^{-}$ colliders.

\begin{figure}[h]
    \centering
\includegraphics[width=8cm]{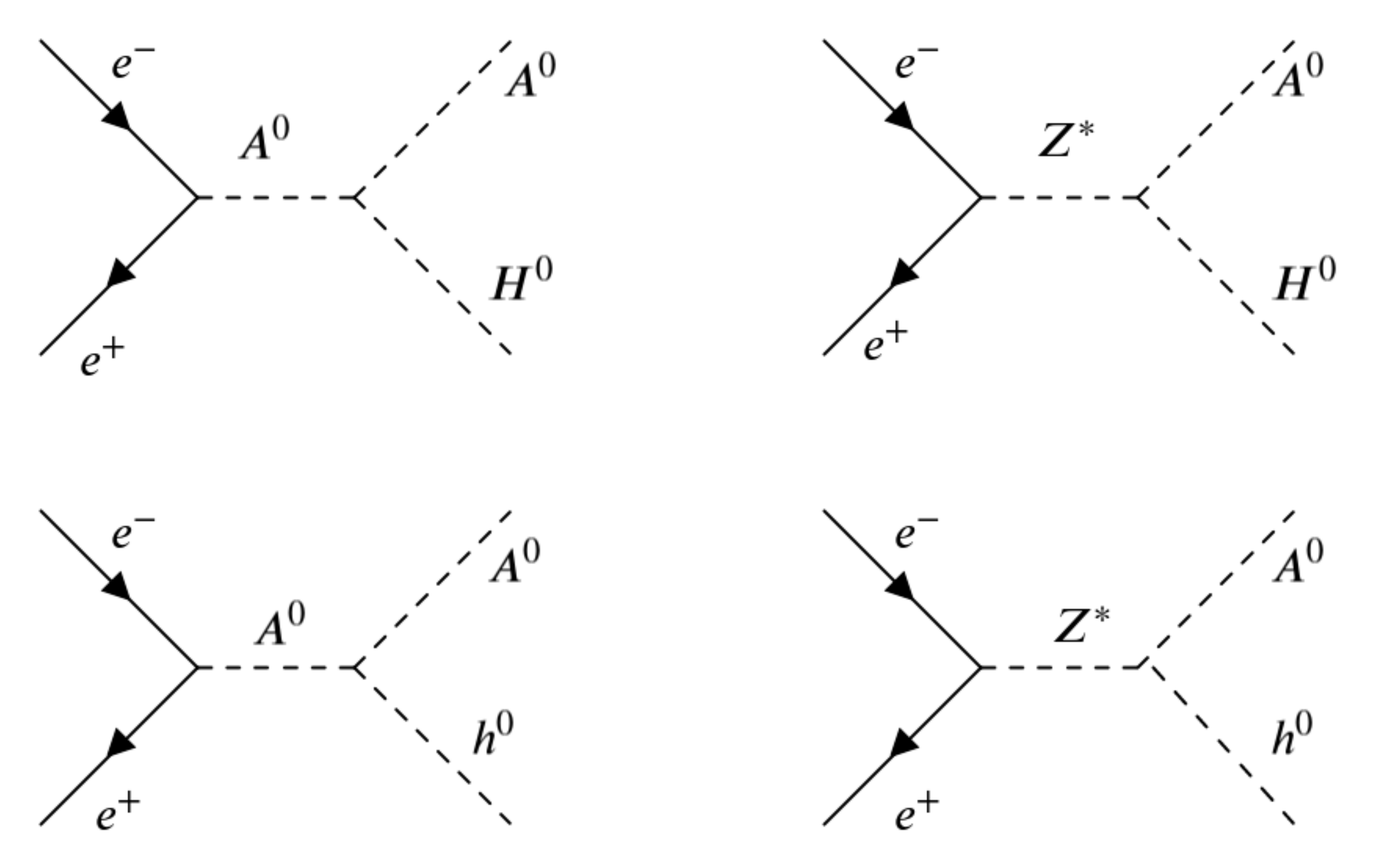}
\caption{\label{fig3} The possible Feynman Diagram for Neutral Higgs pair production}
\end{figure}
These Feynman diagrams are basically $\mu^{+}$ $\mu^{-}$ annihilation diagrams. As a result a mediator particle is produced that further decays into a pair of neutral Higgs A${}^{0}$H${}^{0}$ as well as $A^{0} h^{0}$.

As the table \ref{tab4} and \ref{tab5} contains the value of the assumed benchmark points for  the  processes of neutral Higgs pair productions  $\mu^{+}$ $\mu^{-}$ $\to$ $A^{0}$ $H^{0}$  and $\mu^{+}$ $\mu^{-}$ $\to$ $A^{0}$ $h^{0}$ respectively for various CM energy values at the  future linear $\mu^{+}$$\mu^{-}$  colliders. Here BP is abbreviated as benchmark points. In both the tables there are various parameters $m_{h}$, $m_{H}$, $m_{H-}$, and $m_{A}$ represents the values of Higgs bosons physical masses. As all these values of assumed benchmark points and their corresponding cross-section values are computed at $\tan{\beta}$ =10. Similarly the term of mixing angle in this particular case is $\sin{\beta-\alpha}$ =1. The parameter $m_{{\rm 12}}^{{\rm 2}} $ is basically range which satisfies the theoretical types of requirements. The corresponding signal cross-section values are calculated for each scenario in fb.

\begin{center}
  \begin{table*}[ht]
     \centering	
\begin{tabular}{|c|c|c|c|c|c|c|}  \hline 
 & \textbf{BP1} & \textbf{BP2} & \textbf{BP3} & \textbf{BP4} & \textbf{BP5} & \textbf{BP6} \\  \hline 
$m_{h}$(GeV) & 125 & 125 & 125 & 125 & 125 & 125 \\  \hline 
$m_{H\ }$(GeV) & 150 & 150 & 200 & 200 & 250 & 250 \\  \hline 
$m_{A}$(GeV) & 200 & 250 & 250 & 300 & 300 & 330 \\ \hline  
$m_{H^{\pm}}$(GeV) & 200 & 250 & 250 & 300 & 300 & 330 \\ \hline   
$m_{\rm 12}^{\rm 2} $(GeV)${}^{2}$ & 1987-2243 & 1987-2243 & 3720-3975 & 3720-3975 & 5948-6203 & 5948-6203 \\  \hline 
$\tan{\beta}$ & 10 & 10 & 10 & 10 & 10 & 10 \\  \hline 
$\sin{\beta-\alpha}$ & 1 & 1 & 1 & 1 & 1 & 1 \\  \hline 
$\sigma$ (fb) at $\sqrt{s} $ = 500 GeV & 18.721 & 10.602 & 4.250 & 1.298 & 1.278 & 1.264 \\  \hline 
$\sigma$ (fb) at $\sqrt{s} $ = 1000 GeV & 10.153 & 9.401 & 8.791 & 7.939 & 7.194 & 6.637 \\  \hline 
$\sigma$ (fb) at $\sqrt{s} $ = 1500 GeV & 5.011 & 4.854 & 4.730 & 4.544 & 4.388 & 4.263 \\  \hline 
$\sigma$ (fb) at $\sqrt{s} $ = 2000 GeV & 2.918 & 2.868 & 2.822 & 2.767 & 2.717 & 2.676 \\  \hline 
$\sigma$ (fb) at $\sqrt{s} $ = 2500 GeV & 1.897 & 1.887 & 1.861 & 1.835 & 1.814 & 1.797 \\  \hline 
$\sigma$ (fb) at $\sqrt{s} $ = 3000 GeV & 1.329 & 1.319 & 1.301 & 1.298 & 1.288 & 1.280 \\  \hline 
  \end{tabular}
      \caption{ \label{tab4} Assumed benchmark points and their corresponding cross-sections in fb ($\mu^{+}$ $\mu^{-}$ $\to$ $A^{0}$ $H^{0}$) for various CM energies.}
  \end{table*}
\end{center}

\begin{center}
  \begin{table*}[ht]
     \centering	
\begin{tabular}{|c|c|c|c|c|c|c|}  \hline 
 & \textbf{BP1} & \textbf{BP2} & \textbf{BP3} & \textbf{BP4} & \textbf{BP5} & \textbf{BP6} \\  \hline 
$m_{h}$(GeV) & 125 & 125 & 125 & 125 & 125 & 125 \\  \hline 
$m_{H}$(GeV) & 150 & 150 & 200 & 200 & 250 & 250 \\  \hline 
$m_{A}$(GeV) & 200 & 250 & 250 & 300 & 300 & 330 \\  \hline 
$m_{H^{\pm}}$(GeV) & 200 & 250 & 250 & 300 & 300 & 330 \\ \hline  
$m_{{\rm 12}}^{{\rm 2}} $(GeV)${}^{2}$ & 1987-2243 & 1987-2243 & 3720-3975 & 3720-3975 & 5948-6203 & 5948-6203 \\ \hline  
$\tan{\beta}$ & 10 & 10 & 10 & 10 & 10 & 10 \\  \hline 
$\sin{\beta - \alpha}$) & 1 & 1 & 1 & 1 & 1 & 1 \\  \hline 
$\sigma$ (fb) at $\sqrt{s} $ = 500 GeV & 22.132 & 14.890 & 12.980 & 5.985 & 5.017 & 2.821 \\  \hline 
$\sigma$ (fb) at $\sqrt{s} $ = 1000 GeV & 10.398 & 9.961 & 8.998 & 2.992 & 8.776 & 8.208 \\  \hline 
$\sigma$ (fb) at $\sqrt{s} $ = 1500 GeV & 5.010 & 4.894 & 4.893 & 1.805 & 4.717 & 4.591 \\  \hline 
$\sigma$ (fb) at $\sqrt{s} $ = 2000 GeV & 2.934 & 2.909 & 2.891 & 2.840 & 2.823 & 2.781 \\  \hline 
$\sigma$ (fb) at $\sqrt{s} $ = 2500 GeV & 1.904 & 1.901 & 1.899 & 1.876 & 1.858 & 1.840 \\  \hline 
$\sigma$ (fb) at $\sqrt{s} $ = 3000 GeV & 1.348 & 1.334 & 1.324 & 1.311 & 1.309 & 1.301 \\  \hline 
  \end{tabular}
    \caption{ \label{tab5} Assumed benchmark points and their corresponding cross-sections in fb ($\mu^{+}$ $\mu^{-}$ $\mathrm{\to}$ A${}^{0}$h${}^{0}$) for various CM energies}
  \end{table*}
\end{center}

In both the tables \ref{tab4} and \ref{tab5} there are total six benchmark points (BP)  from BP1 to BP6 which are calculated with their corresponding values of cross-section in fb at specific values of various parameters mentioned above. These values are calculated at CM energy from 500 GeV to 3000 GeV, for BP1 to BP6. As it clear from both the above tables that values for the benchmark points in each case from BP1 to BP6 are greater at CM energy 500 GeV but gradually decreases with increase of CM energy values, hence shows minimum value at 3000 GeV. Similarly in ~both the processes of neutral Higgs pair productions $\mu^{+}$ $\mu^{-}$ $\mathrm{\to}$ A${}^{0}$H${}^{0}$ as well as $\mu^{+}$ $\mu^{-}$ $\mathrm{\to}$ A${}^{0}$h${}^{0}$ the values of benchmark points from BP1 to BP6 at one specific CM energy cross-section values in fb gradually decrease as listed above. Hence the whole conclusion is that cross-section values for every benchmark point shows decreasing behavior on increasing CM energy. 

\begin{figure}[h]
    \centering
\includegraphics[width=8cm]{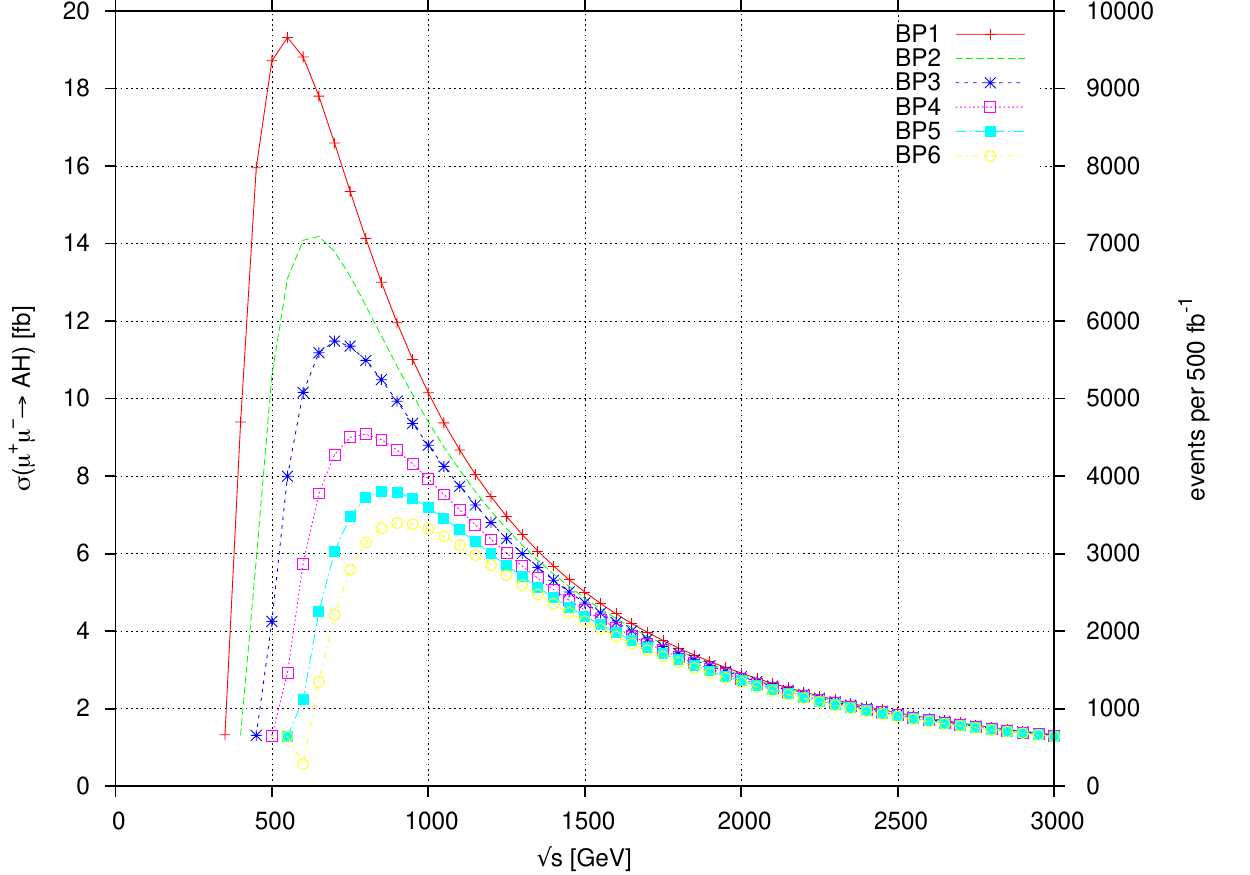}
\caption{\label{fig4} Neutral Higgs pair production ($A^{0}$ $H^{0}$) in 2HDM type-I for various values of $\sqrt{s} $}
\end{figure}

\begin{figure}[h]
    \centering
\includegraphics[width=8cm]{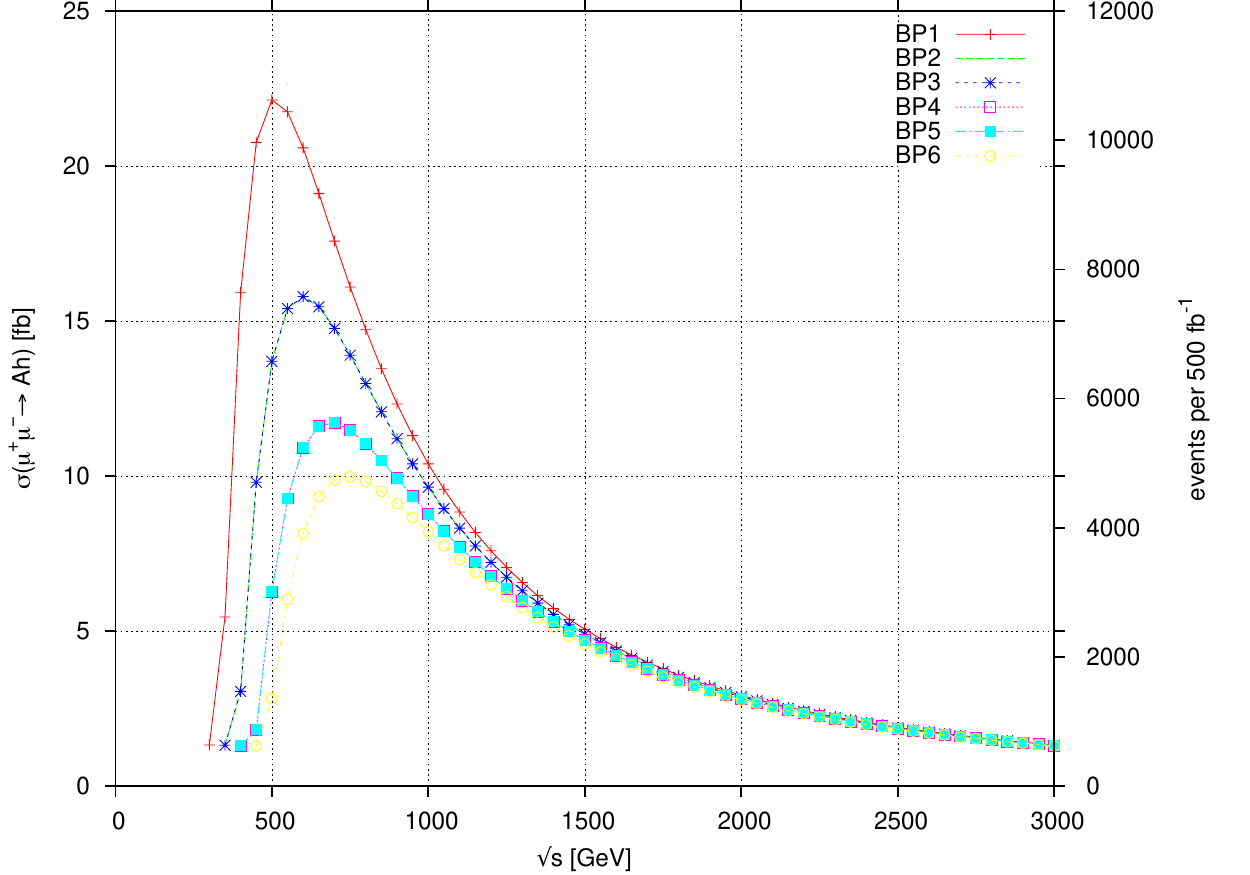}
\caption{Neutral Higgs pair production ($A^{0}$ $h^{0}$) in 2HDM type-I for various values of $\sqrt{s} $ }
\label{fig5} 
\end{figure}

The figure \ref{fig4} and \ref{fig5} are basically graphical forms of table 4 and 5 both the plots shows the benchmark points along with their corresponding total cross-section values in fb and also total number of events per 500 fb${}^{-1}$ in general 2HDM \cite{lab20,lab21}  type-I for neutral higgs pair production processes $\mu^{+}$ $\mu^{-\ }$$\mathrm{\to}$ A${}^{0}$H${}^{0\ }$as well as $\mu^{+}$ $\mu^{-}$ $\mathrm{\to}$ A${}^{0}$h${}^{0}$ at proposed linear muon anti-muon colliders. 

As these curves indicates cross-section values which are function of CM energy values. There are total six curves in both the graphs representing the benchmark points from BP1 to BP6 which contains the corresponding cross-section values in fb at various CM energy values from 500 GeV to 3000 GeV. As all curves shows higher cross-section values and number of events at CM energy of 500 GeV, but shows smaller values of cross-section and number of events at higher CM energy values. So the overall conclusion of above discussion is that in both the types of neutral higgs pair production $\mu^{+}$ $\mu^{-}$ $\to$ A${}^{0}$H${}^{0}$ as well as $\mu^{+}$ $\mu^{-}$ $\to$ $A^{0}$ $h^{0}$ all the curves from BP1 to BP6 are more prominent and also more separated from each other showing their maximum cross-section values at CM energy of 500 GeV, but by increasing CM energy values these curves of benchmark points showing cross-section values become very close to each other representing decreasing behavior at higher CM energy values. Hence at 3000 GeV each curve of benchmark point show their corresponding cross-section values very close to each other as it is depicted by both the above plots. Moreover these curves in both the cases merge into each other showing lower cross-section values and number of events per 500 $fb^{-1}$ at CM energy of 3000 GeV.

Similarly another important observation that has been noticed from Both the processes of neutral Higgs pair productions $\mu^{+}$ $\mu^{-}$ $\to$ $A^{0}$ $H^{0\ }$ as well as $\mu^{+}$ $\mu^{-}$ $\to$ $A^{0}$ $h^{0}$ is that the benchmark points and their corresponding cross-section values shows enhancement for $\mu^{+}$ $\mu^{-}$ $\to$ $A^{0}$ $h^{0}$ as compared to that of $\mu^{+}$ $\mu^{-}$ $\to$ $A^{0}$ $H^{0}$. So both the above tables and graphs shows that cross-section values in fb for neutral Higgs ~production $A^{0}$ $h^{0}$ are higher in magnitude comparatively to that of neutral Higgs production $A^{0}$ $H^{0}$. So the benchmark points and their corresponding cross-section values of neutral Higgs pairs at a specific CM energy are inversely related to mass of neutral Higgs produced in pairs at $\mu^{+}$ $\mu^{-}$ colliders.

\section{ Triple Higgs Production at Future Linear Colliders}
 The triple Higgs (3H) production is analyzed in 2HDM \cite{lab20,lab21}  as well as its special type-II (MSSM) at the proposed linear colliders. The production of (3H) which is considered for study includes various processes such as production of ($H^{+}$, $H^{-}$, $H$), ($H^{+}$, $H^{-}$, $h$) and ($h^{0}$ $h^{0}$ $A^{0}$). the $1^{st}$ processes are analyzed at two different CM energies (1.5 TeV and 3 TeV) in 2HDM as well in MSSM, \cite{lab23,lab24,lab25}  while neutral triple Higgs production ($h^{0}$ $h^{0}$ $A^{0}$) is considered in 2HDM type-I at various CM energy values from 0.5 TeVGeV to 3 TeV. The triple charged Higgs production is analyzed from various papers in the context of future linear colliders \cite{lab27,lab28,lab29,lab30,lab31,lab32} . As a process of triple Higgs production ($H^{+}$ $H^{-}$ $H^{0}$) as well as ($H^{+}$ $H^{-}$ $h^{0}$) is under consideration in the annihilation process of $\mu^{+}$$\mu^{-}$, within the type II of 2HDM \cite{lab20,lab21} . As their tri-linear couplings of Higgs bosons are shown by equation (\ref{eq7} and \ref{eq8}).

\begin{center}
\begin{equation}  
  \label{eq7}
H^{\pm} H^{\pm} H^{0} \quad (2HDM)  :  \frac{-e}{m_{W} {\rm .}s_{W} {\rm .}s_{{\rm 2}\beta }^{{\rm 2}} } \left(c_{\beta }^{{\rm 3}} s_{{\rm 2}\beta } s_{\alpha } m_{H}^{{\rm 2}} +c_{\alpha } s_{{\rm 2}\beta } s_{\beta }^{{\rm 3}} m_{H}^{{\rm 2}} -{\rm 2}s_{\alpha +\beta } \mu _{{\rm 12}}^{{\rm 2}} +c_{\beta -\alpha } s_{{\rm 2}\beta }^{{\rm 2}} m_{H^{\pm } }^{{\rm 2}} \right)
\end{equation}
\begin{equation}  
  \label{eq8}
 H^{\pm} H^{\pm} h^{0}  \quad (2HDM)  :  \frac{e}{m_{W} {\rm .}s_{W} {\rm .}s_{{\rm 2}\beta }^{{\rm 2}} } \left({\rm 2}c_{\alpha +\beta } \mu _{{\rm 12}}^{{\rm 2}} -c_{\alpha } c_{{\rm 2}\beta }^{{\rm 3}} m_{h}^{{\rm 2}} +s_{2\beta } s_{\alpha } s_{\beta }^{{\rm 3}} m_{h}^{{\rm 2}} -s_{{\rm 2}\beta }^{{\rm 2}} s_{\beta -\alpha } m_{H^{\pm } }^{{\rm 2}} \right) 
\end{equation}
\begin{equation} 
  \label{eq9} 
 H^{\pm} H^{\pm} H^{0}  \quad (MSSM)  :  \frac{-e{\rm .}m_{Z}^{{\rm 2}} }{{\rm 2}m_{W} {\rm .}s_{W} } \left(c_{{\rm 2}W} c_{\beta -\alpha } +s_{2\beta } s_{\alpha +\beta } \right)
\end{equation}
\begin{equation}
  \label{eq10}
H^{\pm} H^{\pm} h^{0}  \quad (MSSM)  :  \frac{-e{\rm .}m_{Z}^{{\rm 2}} }{{\rm 2}m_{W} {\rm .}s_{W} } \left({\rm 2}c_{W}^{{\rm 2}} s_{\beta -\alpha } +c_{2\beta } s_{\alpha +\beta } \right)
\end{equation}
\end{center}

\begin{figure}[h]
    \centering
\includegraphics[width=8cm]{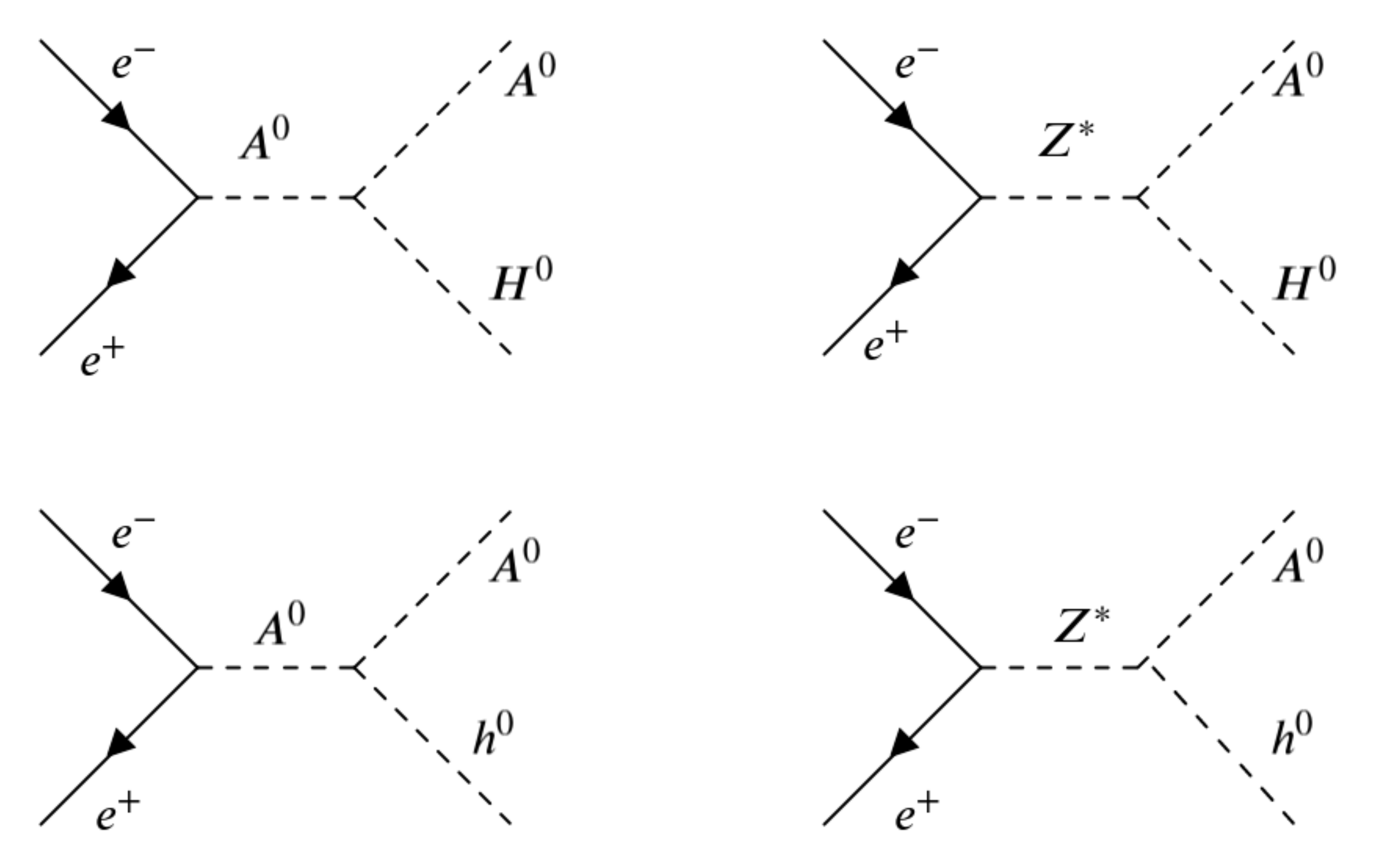}
\caption{\label{fig6} The possible Feynman Diagrams for Triple Higgs production at lepton collider.}
\end{figure}

 Similarly, their MSSM \cite{lab23,lab24,lab25}  corresponding values are in equation (\ref{eq9} and \ref{eq10}). Here some specific abbreviations just like $s_{W} = \sin{\theta_{W}}$ and $s_{\beta} = \sin{\beta}$ are used \cite{lab51}  as figure \ref{fig6} depicts the production of (H${}^{+}$H${}^{-}$H${}^{0}$). The couplings of these charged Higgs efficaciously increase with $\tan{\beta}$ $\mathrm{>}$$\mathrm{>}$1 or with $\cot{\beta}$ for $\tan{\beta}$ $\mathrm{<}$$\mathrm{<}$ 1. as the values of the cross section can change either by $\tan{2\beta}$ at higher values of $\tan{\beta}$ or by $\cot{2\beta}$ at lower values of $\tan{\beta}$ respectively in comparison with the case in MSSM. Here the couplings of triple Higgs goes through radiative corrections \cite{lab51} and have not any kind of enhancing source as mentioned by equation \ref{eq10}. This equation clearly described that there is a natural type of gauge couplings, so the corrections one can expect becomes smaller. Figure \ref{fig6} represents the possible Feynman diagrams at the lepton colliders for triple Higgs production.

Table \ref{tab6} and \ref{tab7} contains the value of cross-section for triple Higgs production. There are two types of triple Higgs production which are studied at linear muon anti muon collider, they are included (H${}^{+}$,H${}^{-}$,h${}^{0}$) and (H${}^{+}$,H${}^{-}$,H${}^{0}$). The cross-section values are calculated under 2HDM \cite{lab20,lab21}  and its special type-II (MSSM) at two different CM energies. 

\begin{center}
  \begin{table*}[ht]
     \centering	
\begin{tabular}{|c|c|c|c|c|c|}  \hline 
Process & $m_{H^{\pm}}$ GeV & $\sigma_{max}$ (1.5TeV) MSSM & $\sigma_{max}$ (3TeV) MSSM & $\sigma_{max}$ (1.5TeV) 2HDM & $\sigma_{max}$ (3TeV) 2HDM \\  \hline 
$\mu^{+}$ $\mu^{-}$ $\to$ $H^{+}$ $H^{-}$ $h^{0}$ & 500 & 1.0523x10${}^{-5}$ & 1.2130x10${}^{-5}$ & 3.7347x10${}^{-4}$ & 4.4302x10${}^{-4}$ \\  \hline 
$\mu^{+}$ $\mu^{-}$ $\to$ $H^{+}$ $H^{-}$ $h^{0}$ & 600 & 3.4356x10${}^{-6}$ & 1.1030x10${}^{-5}$ & 5.6166x10${}^{-5}$ & 3.3321x10${}^{-4}$ \\  \hline 
$\mu^{+}$ $\mu^{-}$ $\to$ $H^{+}$ $H^{-}$ $h^{0}$ & 700 & NP & 9.1982x10${}^{-6}$ & NP & 2.2234x10${}^{-4}$ \\  \hline 
$\mu^{+}$ $\mu^{-}$ $\to$ $H^{+}$ $H^{-}$ $h^{0}$ & 800 & NP & 6.8499x10${}^{-6}$ & NP & 1.1952x10${}^{-4}$ \\  \hline 
$\mu^{+}$ $\mu^{-}$ $\to$ $H^{+}$ $H^{-}$ $h^{0}$ & 900 & NP & 4.8158x10${}^{-6}$ & NP & 1.0153 x10${}^{-4}$ \\  \hline 
$\mu^{+}$ $\mu^{-}$ $\to$ $H^{+}$ $H^{-}$ $h^{0}$ & 1000 & NP & 3.1719x10${}^{-6}$ & NP & 0.0101 x10${}^{-4}$\newline  \\  \hline 
  \end{tabular}
    \caption{ \label{tab6} Total cross-section in (fb) of Triple Higgs ($H^{+}$ $H^{-}$ $h^{0\ }$)at $\sqrt{s} $ =1.5TeV \& $\sqrt{s} $=3TeV}
  \end{table*}
\end{center}

\begin{center}
  \begin{table*}[ht]
     \centering	
\begin{tabular}{|c|c|c|c|c|c|}  \hline 
Process & $m_{H^\pm}$ GeV & $\sigma_{max}$ (1.5TeV) MSSM & $\sigma_{max}$ (3TeV) MSSM & $\sigma_{max}$ (1.5TeV) 2HDM & $\sigma_{max}$ (3TeV) 2HDM \\  \hline 
$\mu^{+} \mu^{-} \to H^{+} H^{-} H^{0}$ & 500 & NP & $3.6801 x10^{-6}$ & NP & $0.12$ \\  
$\mu^{+} \mu^{-} \to H^{+} H^{-} H^{0}$ & 600 & NP & $1.5011 x10^{-6}$ & NP & $3.1241x10^{-2}$ \\  
$\mu^{+} \mu^{-} \to H^{+} H^{-} H^{0}$ & 700 & NP & $3.86 16x10^{-7}$ & NP & $1.8101 x10^{-2}$ \\  
$\mu^{+} \mu^{-} \to H^{+} H^{-} H^{0}$ & 800 & NP & $1.6211 x10^{-7}$ & NP & $6.4302 x10^{-3}$ \\  
$\mu^{+} \mu^{-} \to H^{+} H^{-} H^{0}$ & 900 & NP & $3.8601 x10^{-8}$ & NP & $5.7771 x10^{-4}$ \\ \hline  
  \end{tabular}
    \caption{ \label{tab7} Total cross-section in (fb) of Triple Higgs at ($H^{+}$ $H^{-}$ $H^{0}$ ) $\sqrt{s} $ =1.5TeV \& $\sqrt{s} $=3TeV}
  \end{table*}
\end{center}

\begin{figure}[h]
    \centering
\includegraphics[width=8cm]{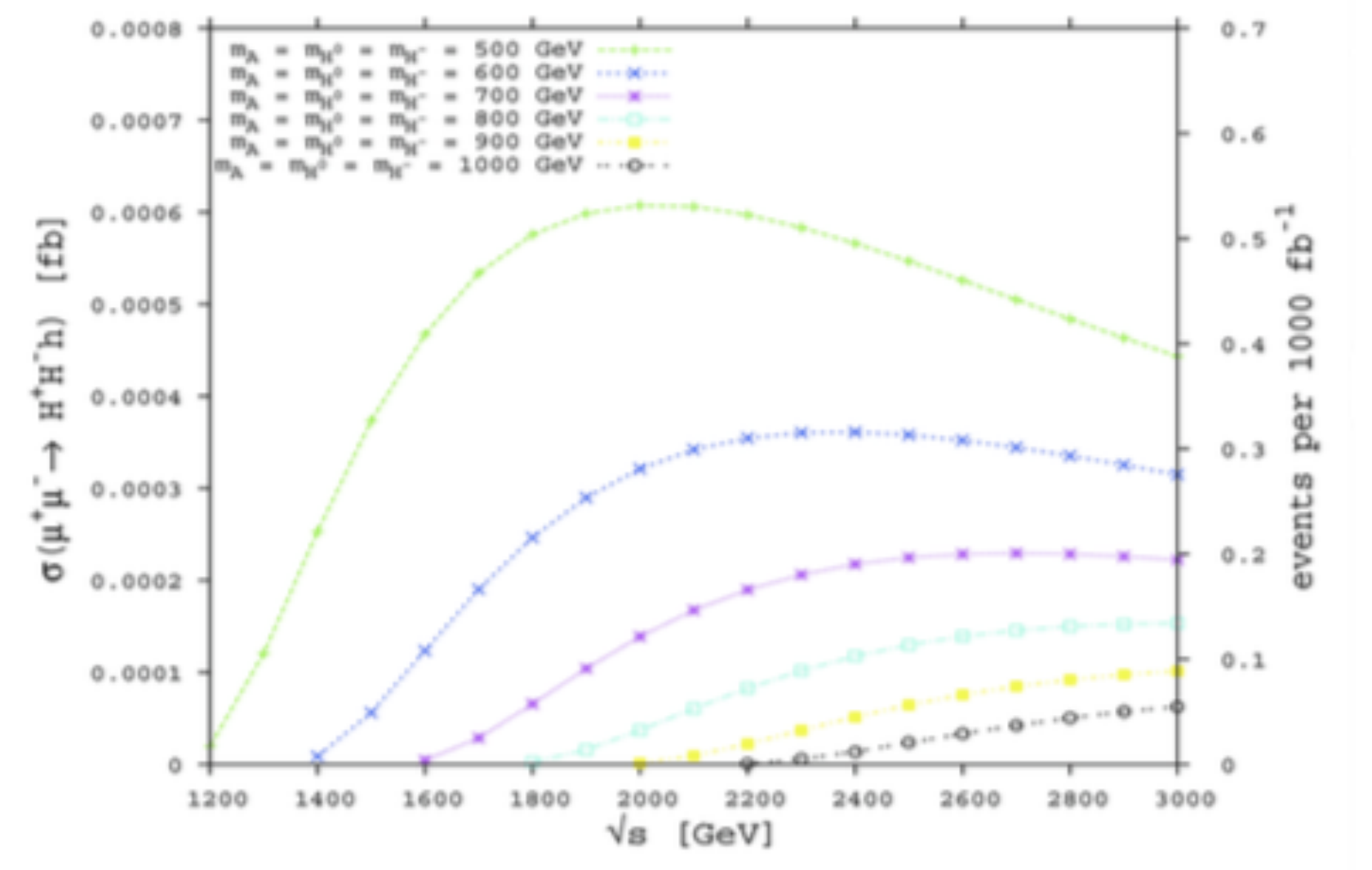}
\caption{\label{fig7} Triple Higgs (H${}^{+}$H${}^{-}$h${}^{0}$) production in 2HDM type-II at various values of m${}_{H}$$\mathrm{\pm}$}
\includegraphics[width=8cm]{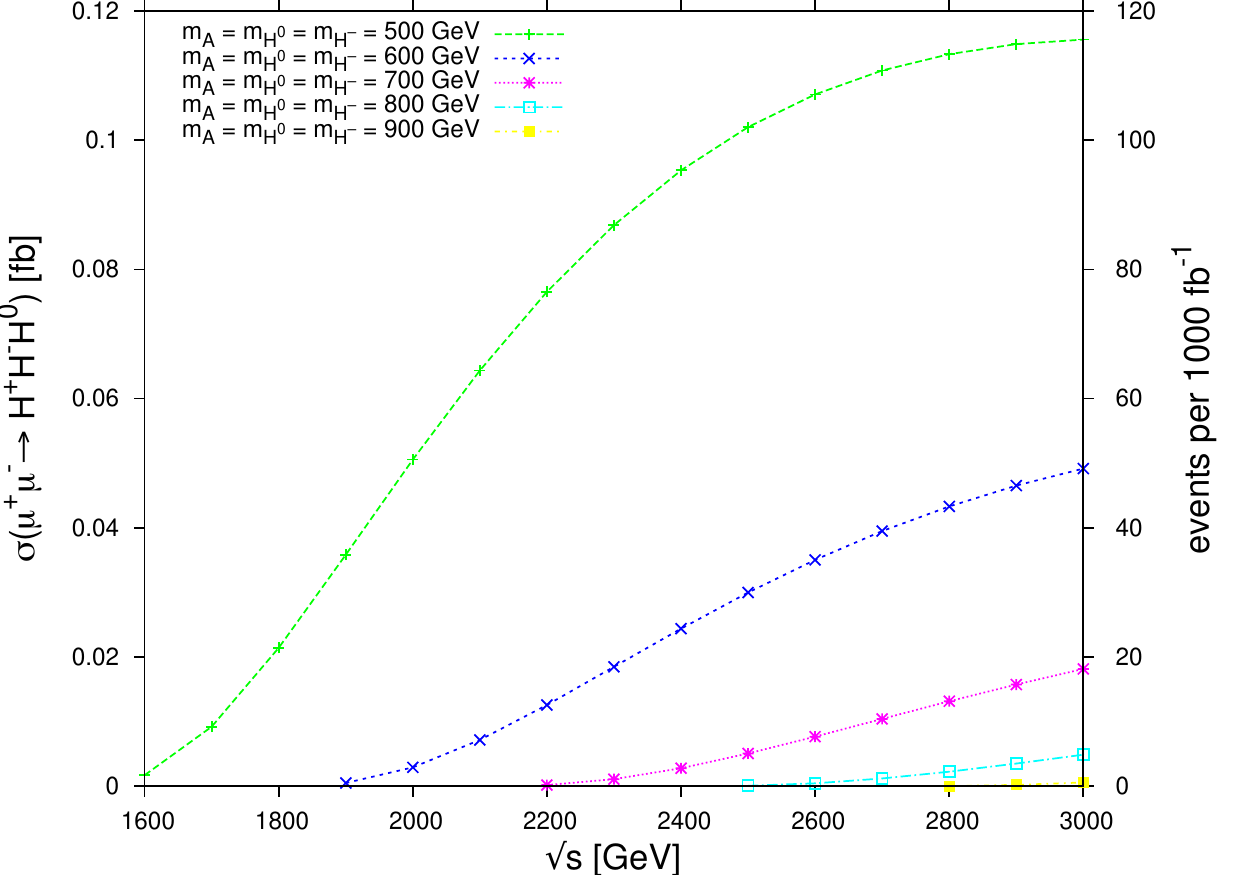}
\caption{\label{fig8} Triple Higgs (H${}^{+}$H${}^{-}$H${}^{0}$) production in 2HDM type-II at various values of m${}_{H}$$\mathrm{\pm}$}
\end{figure}

As it is clear from above tables the values of cross-section in fb for both the kind of processes are inversely related with mass of charged Higgs \cite{lab18}  . It is concluded from above all discussion that cross-section values are decreased with the increase of masses of charged Higgs. Figure \ref{fig7} and \ref{fig8} represents triple charged Higgs productions (H${}^{+\ }$H ${}^{-}$h${}^{0}$)  as well as (H${}^{+\ }$H${}^{-\ }$H${}^{0}$)  at various CM energies. These graphs in figure  \ref{fig7} and \ref{fig8}  represents maximum values of cross-section in fb as a function of center of mass energy for triple Higgs production (H${}^{+}$H${}^{-}$h${}^{0}$)  as well as (H${}^{+}$H${}^{-}$H${}^{0}$) respectively. These curves in both the graphs  \ref{fig7} and \ref{fig8}  depict the values of cross-section in fb at different values of CM energies and mass of the charged Higgs.

In both these cases the mass of sudo-scalar Higgs (CP odd Higgs A), mass of the heavy Higgs (CP even Higgs H${}^{0}$) and the mass of charged higgs is equal as listed below. Similarly the value of most important factor $\tan{\beta}$ is 10 and the light Higgs mass (h) is 125 GeV. As the cross-section values are studied at various values of charged Higgs mass \cite{lab18} . In both these graphs it be seen that when the mass of charged Higgs is smaller about 500GeV then a curve having higher values of cross-section at different values of center of mass energy is obtained. Similarly in other case, when the mass of charged Higgs is increased up to 900GeV or 1000GeV, and all the other factor remains constant then a curve of lower values is obtained in both the cases indicating the values of cross-section at corresponding values of CM energies. Similarly the number of events per 1000fb${}^{-1\ }$(luminosities) values also decreases for higher values of charged Higgs mass. So cross-section is inversely related to mass of charged Higgs \cite{lab18}  .

\begin{center}
  \begin{table*}[ht]
     \centering	
\begin{tabular}{|c|c|c|c|c|c|c|c|}  \hline 
\textbf{Type of Collider} & \textbf{Type of Model} & \textbf{Z/$\boldsymbol{\gamma}$} & \textbf{H${}^{+}$H${}^{-}$H${}^{0}$\newline h${}^{0}$/H${}^{0}$} & \textbf{Total} & \textbf{Z/$\boldsymbol{\gamma}$} & \textbf{H${}^{+}$H${}^{-}$h${}^{0}$\newline h${}^{0}$/H${}^{0}$} & \textbf{Total} \\  \hline 
$\mu^{+}$$\mu^{-}$ & MSSM & 1.92x10${}^{-7}$ & 3.48x10${}^{-6}$ & 3.68x10${}^{-6}$ & 1.68x10${}^{-5}$ & 2.84x10${}^{-8}$ & 1.68x10${}^{-5}$ \\  
$\mu^{+}$$\mu^{-}$ & 2HDM & 0.115 & 2.25x10${}^{-6}$ & 0.115 & 4.43x10${}^{-4}$ & 1.91x10${}^{-10}$ & 4.43x10${}^{-4}$ \\  
e${}^{+}$ e${}^{-}$ & MSSM & 1.92x10${}^{-7}$ & 7.76x10${}^{-11}$ & 1.92x10${}^{-7}$ & 1.68x10${}^{-5}$ & 6.35x10${}^{-13}$ & 1.68x10${}^{-5}$ \\  
e${}^{+}$ e${}^{-}$ & 2HDM & 0.115 & 5.11x10${}^{-11}$ & 0.115 & 4.43x10${}^{-4}$ & 1.02x10${}^{-19}$ & 4.43x10${}^{-4}$ \\  \hline 
  \end{tabular}
    \caption{ \label{tab8} Total cross-section in (fb) of triple Higgs at $\sqrt{s} $ =1.5TeV under MSSM \& 2HDM}
  \end{table*}
\end{center}

Table \ref{tab8} shows in case of triple Higgs production an enhancement of several orders of about ($\mathrm{\sim}$ $10^{4}$) in magnitude is seen in 2HDM as compared to its special type-II MSSM. It is because of greater coupling of charged Higgs Bosons in 2HDM as compared to MSSM \cite{lab23,lab24,lab25}  as described in equation \ref{eq7} and \ref{eq8} for this reason 2HDM is preferred for calculating the cross-section of charged Higgs \cite{lab18} .

\subsection{ Cross-Section Comparison in MSSM and 2HDM for Triple Higgs Production}
 Cross-Section Comparison is made in MSSM \cite{lab23,lab24,lab25}  and 2HDM \cite{lab20,lab21}  for Triple Higgs Production. The overall values of cross-section in fb for both the models are listed in table \ref{tab9}.

\begin{center}
  \begin{table*}[ht]
     \centering	
\begin{tabular}{|c|c|c|c|c|c|c|c|}  \hline 
\textbf{Type of Collider} & \textbf{Type of Model} & \textbf{Z/$\boldsymbol{\gamma}$} & \textbf{H${}^{+}$H${}^{-}$H${}^{0}$\newline h${}^{0}$/H${}^{0}$} & \textbf{Total} & \textbf{Z/$\boldsymbol{\gamma}$} & \textbf{H${}^{+}$H${}^{-}$h${}^{0}$\newline h${}^{0}$/H${}^{0}$} & \textbf{Total} \\  \hline 
$\mu^{+}$$\mu^{-}$ & MSSM & 1.92x10${}^{-7}$ & 3.48x10${}^{-6}$ & 3.68x10${}^{-6}$ & 1.68x10${}^{-5}$ & 2.84x10${}^{-8}$ & 1.68x10${}^{-5}$ \\  \hline 
$\mu^{+}$$\mu^{-}$ & 2HDM & 0.115 & 2.25x10${}^{-6}$ & 0.115 & 4.43x10${}^{-4}$ & 1.91x10${}^{-10}$ & 4.43x10${}^{-4}$ \\  \hline 
e${}^{+}$ e${}^{-}$ & MSSM & 1.92x10${}^{-7}$ & 7.76x10${}^{-11}$ & 1.92x10${}^{-7}$ & 1.68x10${}^{-5}$ & 6.35x10${}^{-13}$ & 1.68x10${}^{-5}$ \\  \hline 
e${}^{+}$ e${}^{-}$ & 2HDM & 0.115 & 5.11x10${}^{-11}$ & 0.115 & 4.43x10${}^{-4}$ & 1.02x10${}^{-19}$ & 4.43x10${}^{-4}$ \\  \hline 
  \end{tabular}
    \caption{ \label{tab9} Total cross-section in (fb) of triple Higgs at $\sqrt{s} $ = 1.5TeV under MSSM \& 2HDM}
  \end{table*}
\end{center}

Table \ref{tab8} shows in case of triple Higgs production an enhancement of several orders of about ($\mathrm{\sim}$10${}^{4}$) in magnitude is seen in 2HDM as compared to its special type-II MSSM. It is because of greater coupling of charged Higgs Bosons in 2HDM as compared to MSSM \cite{lab23,lab24,lab25}  as described in equation \ref{eq7} and \ref{eq8} for this reason 2HDM is preferred for calculating the cross-section of charged Higgs \cite{lab18} .

\section{Neutral Triple Higgs Production}
 In neutral triple Higgs production ($\mu^{+} \mu^{-} \to h^{0} h^{0} A^{0}$) process is considered for study in the framework of 2HDM type-I. This process of neutral  triple Higgs production is studied in the context of most significant future linear colliders \cite{lab27,lab28,lab29,lab30,lab31},  at various CM energies from 500 GeV to 3000 GeV. This process of neutral triple Higgs production $\mu^{+} \mu^{- } \to h^{0} h^{0} A^{0}$ is studied at proposed linear muon anti-muon ($\mu^{+}$ $\mu^{-}$) colliders \cite{lab13,lab14,lab15} . The main purpose of this analysis is to understand the neutral triple Higgs production process and also the sources of neutral Higgs in 2HDM type-I. This theoretical study is done by assuming a number of benchmark points in 2HDM type-I parameter space. Moreover the evidence of predicted pseudo-scalar or CP-odd Higgs (A) and a pair of light or SM Higgs (h) is investigated. The cross-section values in fb are computed at specific values of various parameters. The tri-linear coupling equation is given as \ref{eq11}.

\begin{equation}
\label{eq11}
h^{0} h^{0} A^{0}  \quad (2HDM type-I)  :  \frac{-ie{\rm cos}\left(\beta -\alpha \right)}{{\rm 2}M_{W} {\rm sin}\theta _{W} {\rm sin2}\beta} ( 2M_{h}^{2} +M_{A}^{2} \sin{2 \alpha} - M_{H}^{ 2}( 3\sin{2\alpha} - \sin{2\beta} ) )
\end{equation}

\begin{figure}[h]
    \centering
\includegraphics[width=8cm]{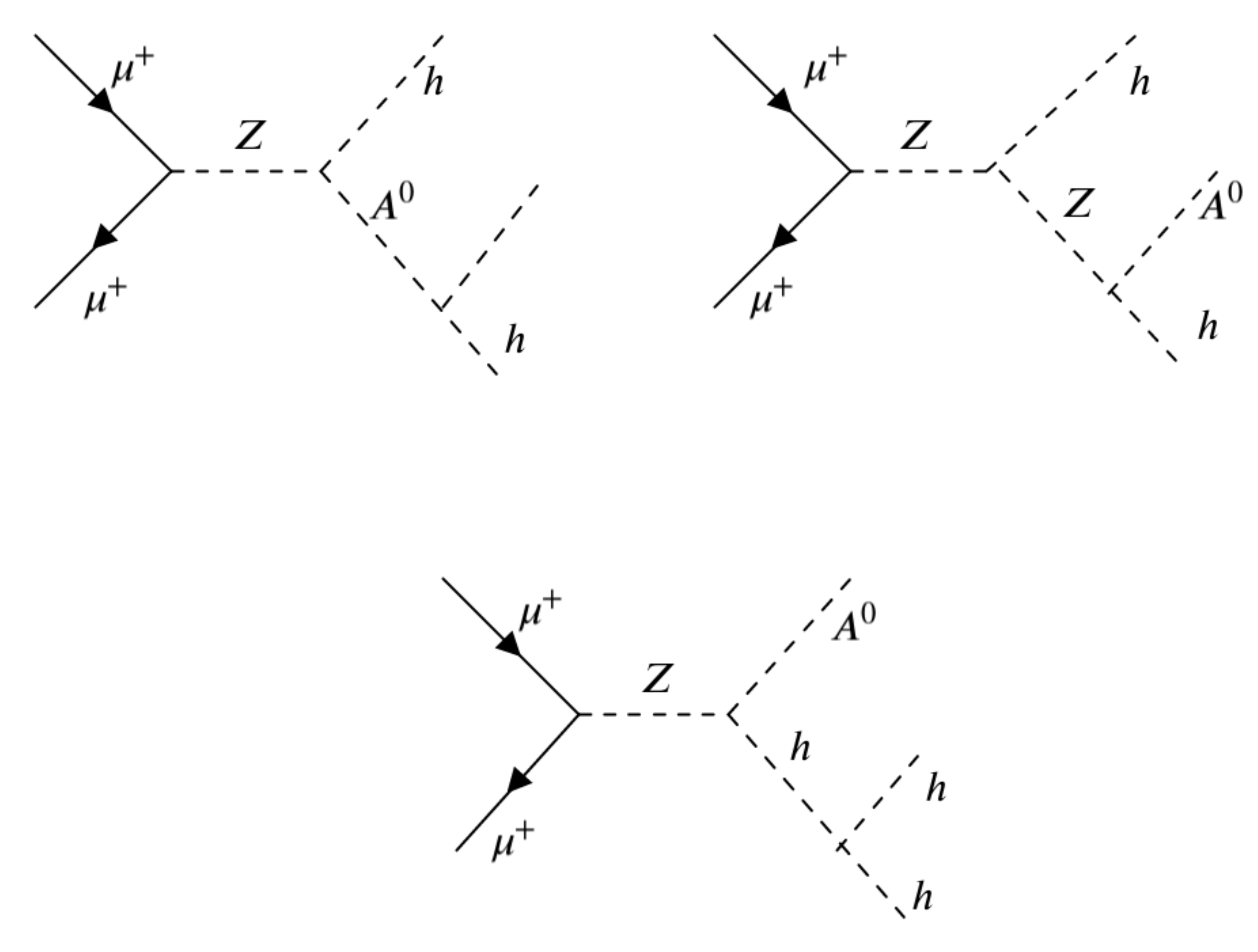}
\caption{\label{fig9} Double Charged Higgs (H${}^{+}$H${}^{-}$) pair production cross-section in (fb) at Liner Muon Colliders under 2HDM or MSSM. }
\end{figure}

The figure \ref{fig9} shows the possible Feynman diagrams of neutral triple Higgs production for $\mu^{+}$ $\mu^{-}$$\mathrm{\to}$ h${}^{0}$h${}^{0}$A${}^{0}$ process at the future linear $\mu^{+}$ $\mu^{-}$ colliders \cite{lab13,lab12,lab13,lab14,lab15} .

The table \ref{tab9} contains the value of assumed benchmark points and their cross-sections in fb at $\tan{\beta}$=20, $\tan{\beta}$=30, $\tan{\beta}$=40 and $\tan{\beta}$=60 for various values of CM energy from 500 GeV to 3000 GeV. 

\begin{center}
  \begin{table*}[ht]
     \centering	
\begin{tabular}{|c|c|c|c|c|}  \hline 
  & \textbf{BP1 tan$\boldsymbol{\beta}$=20} & \textbf{BP1 tan$\boldsymbol{\beta}$=30} & \textbf{BP1 tan$\boldsymbol{\beta}$=40} & \textbf{BP1 tan$\boldsymbol{\beta}$=60} \\ \hline  
m${}_{h }$ (GeV) & 125 & 125 & 125 & 125 \\  \hline 
m${}_{H }$ (GeV) & 150 & 150 & 200 & 200 \\  \hline 
m${}_{A}$ ${}_{\ }$(GeV) & 200 & 250 & 250 & 300 \\  
m${}_{H}$${}^{\mathrm{\pm }\ }$ (GeV) & 200 & 250 & 250 & 300 \\  \hline 
$m_{{\rm 12}}^{{\rm 2}} $(GeV)${}^{2}$ & 1987-2243 & 1987-2243 & 3720-3975 & 3720-3975 \\  \hline 
$\tan{\beta}$ & 10 & 10 & 10 & 10 \\  \hline 
$\sin{\beta-\alpha}$ & 1 & 1 & 1 & 1 \\  \hline 
$\sigma$ (fb) at $\sqrt{s} $ = 500 GeV & 0.743 & 1.678 & 2.986 & 6.723 \\  \hline 
$\sigma$ (fb) at $\sqrt{s} $ = 1000 GeV & 11.111 & 25.069 & 44.610 & 100.44 \\  \hline 
$\sigma$ (fb) at $\sqrt{s} $ = 1500 GeV & 7.175 & 16.119 & 28.811 & 64.871 \\  \hline 
$\sigma$ (fb) at $\sqrt{s} $ = 2000 GeV & 4.566 & 10.307 & 8.341 & 41.279 \\  \hline 
$\sigma$ (fb) at $\sqrt{s} $ = 2500 GeV & 3.075 & 6.939 & 12.350 & 27.806 \\  \hline 
$\sigma$ (fb) at $\sqrt{s} $ = 3000 GeV & 2.202 & 4.969 & 8.842 & 19.910 \\ \hline 
  \end{tabular}
    \caption{ \label{tab10} Assumed benchmark points and their corresponding cross-sections in fb ($\mu^{+}$ $\mu^{-}$ $\mathrm{\to}$ h${}^{0}$h${}^{0}$A${}^{0}$) for various CM energies}
  \end{table*}
\end{center}

It is clear from table 9 that at a specific value of CM energy by increasing $\tan{\beta}$ value the resulting cross-sections in fb shows enhancement. So cross-section of neutral triple Higgs is directly related with the value of $\tan{\beta}$. The couplings of these neutral triple Higgs bosons efficaciously increase with $\tan{\beta}$ $\mathrm{>}$10.

\begin{figure}[h]
    \centering
\includegraphics[width=8cm]{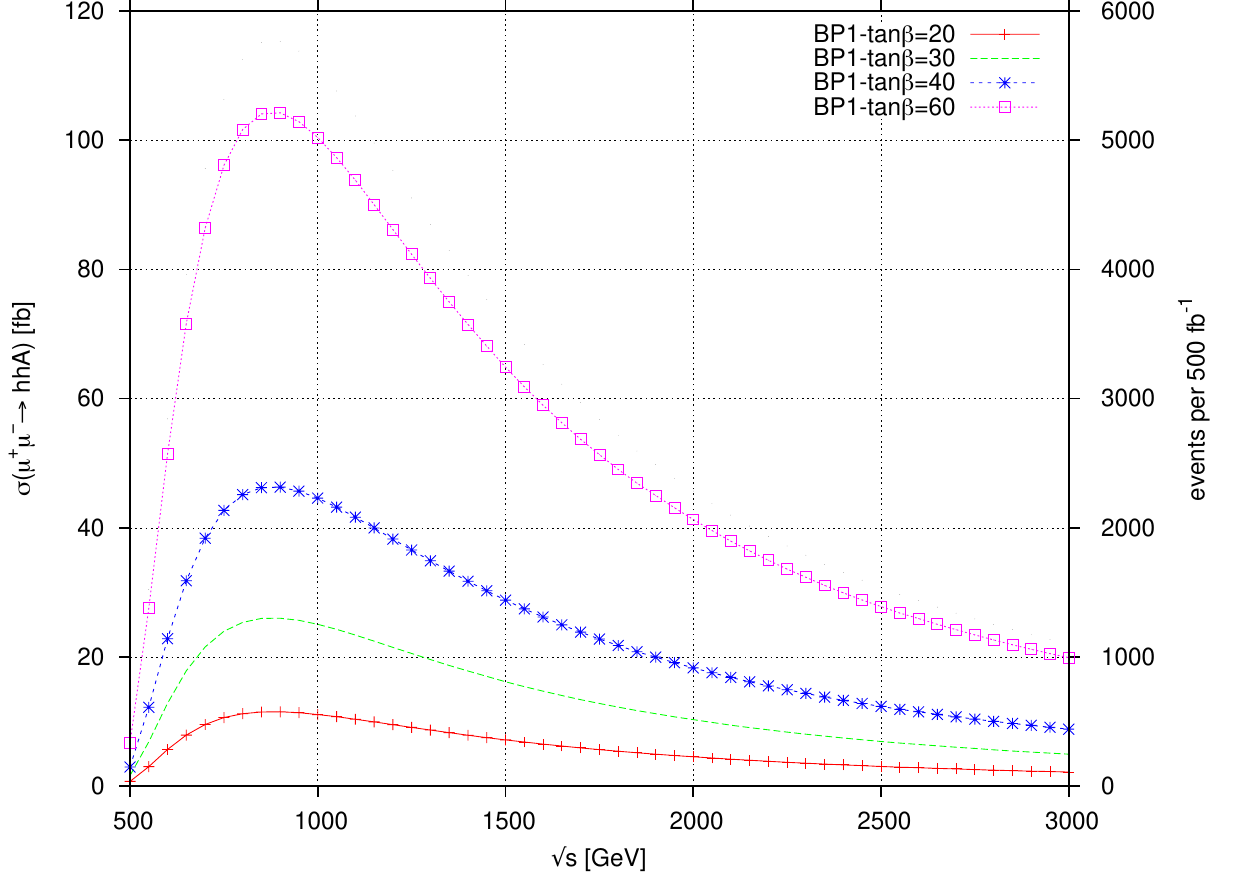}
\caption{\label{fig10} Neutral Triple Higgs production (h${}^{0}$ h${}^{0}$A${}^{0}$) in 2HDM type-I for various values of $\sqrt{s} $  }
\end{figure}

The figure \ref{fig10} is the graphical form of table \ref{tab9}. As there are total four curves representing the benchmark points and corresponding cross-section values computed at $\tan{\beta}$=20, $\tan{\beta}$=30, $\tan{\beta}$=40 and $\tan{\beta}$=60 for various CM energy values from 500 GeV to 3000 GeV.

It is clear from the plot at lower $\tan{\beta}$ value for a specific value of CM energy the resulting cross-section values are smaller but for higher value of $\tan{\beta}$ the resulting cross section values are higher. So at a specific value of CM energy by increasing the $\tan{\beta}$ the resulting cross section show enhanced values so these enhanced values shows that cross-section of neutral triple Higgs (h${}^{0}$h${}^{0}$A${}^{0}$) is directly related to $\tan{\beta}$ value.

\section{DISCUSSION AND CONCLUSIONS}
 In this thesis we analyzed such processes which act as a source of charged Higgs pairs at linear colliders, \cite{lab27,lab28,lab39,lab30,lab31}  at two different center of mass energies i.e. (1.5 TeV and 3 TeV) using two models. 1${}^{st}$ one is two Higgs Doublets Model \cite{lab20,lab21}  and 2${}^{nd}$ one is its special type-II (Minimal Super Symmetric Standard Model) \cite{lab23,lab24,lab25}. Basically MSSM \cite{lab23,lab24,lab25} is a special type-II of 2HDM, and also both these models are extended forms of SM \cite{lab1,lab2,lab3,lab4} . Both these models are extension of SM containing multiple Higgs bosons including charged Higgs ($H^{+}$, $H^{-}$) \cite{lab18}  . For the study of the charged Higgs these models are extensively used by the physicists to understand sources and all phenomena related with charged Higgs. The process of double charged Higgs ($H^{+}$, $H^{-}$)  and triple Higgs production ($H^{+}$, $H^{-}$, $h^{0}$), ($H^{+}$, $H^{-}$, $H^{0}$), is analyzed in the context of proposed future linear colliders \cite{lab27,lab28,lab39,lab30,lab31} in two above models. The main aim is to study and compare the charged Higgs sources via double Higgs as well as triple Higgs production. In both these models 2HDM and its type-II, cross-section is calculated in femtobarn (fb) for various values of charged Higgs masses. As the plot 2 for the double Higgs production represents the cross-section is a function of center of mass energy. It means cross-section depends on the center of mass energy. When the mass of charged Higgs is about 500 GeV then curve shows the higher values of cross-sections at various values of CM energy. However when the mass of charged Higgs is about 1000 Gev then similar curve is obtained but gives lower values of cross-sections at different values of CM energy. Also similar results are obtained in triple Higgs production in graphs 7 and 8 hence proved that cross-section is inversely related with mass of charged Higgs \cite{lab18}  because by increasing mass of charged Higgs values of cross-sections values decreases.

In the process of double charged Higgs production, for a specific value of charged Higgs \cite{lab18}  mass as well as CM energy the maximum values of cross-section almost remains the same in both the models, in 2HDM and its special type-II, but in case of triple Higgs production an enhancement of several orders about ($\sim10^{4}$) in magnitude is seen in 2HDM \cite{lab20,lab21} as compared to its special type-II (MSSM). It is because of greater couplings of charged Higgs bosons in 2HDM as compared in MSSM \cite{lab23,lab24,lab25} according to equations \ref{eq7}, \ref{eq8}, \ref{eq9} and \ref{eq10}. For this reason 2HDM is preferred for calculating the cross-section of charged Higgs.

Another very important conclusion is that values of cross-section are higher in magnitude for double Higgs production as compared to that of cross-section values for triple Higgs production. Thus considering it's an important and main source at linear colliders of charged Higgs. Similarly pair wise Higgs production is very less limited to CM energy of system than that of triple Higgs production. Thus examine heavy charged Higgs \cite{lab18}  at a given value of CM energy.

 Another very significant conclusion is that there is no sizable enhancement is observed in linear electron-positron collider \cite{lab10,lab11,lab12}  versus muon-anti muon collider \cite{lab13,lab14,lab15} . Both these colliders give almost similar values of cross-sections for a specific process at the similar values of CM energies. However there are many advantages of muon-anti muon linear collider over electron-positron collider. It would be able to give more accurate and precise results than electron-positron collider. Similarly it would be best in terms of luminosity and power consumption. The factor of synchrotron radiations in muon-anti muon collider is negligible as compared electron-positron collider. That's why muon-anti muon collider is considered more efficient for calculating the cross-sections as compared to electron-positron collider.

Similarly two types of processes, neutral Higgs pair production ($A^{0} H^{0}$) as well as ($A^{0}$ $h^{0}$) and the triple neutral Higgs production ($h^{0} h^{0} A^{0}$) is analyzed in 2HDM \cite{lab20,lab21}  type-I at the future linear $\mu^{+} \mu^{- }$ collider for various values of center of mass energies from 500 GeV to 3000 GeV. It is observed that the cross-section in fb for neutral Higgs pair production (A${}^{0}$h${}^{0}$) process is comparatively greater than that of neutral Higgs pair production ($A^{0} H^{0}$) process. The reason is that in neutral Higgs pair production process, neutral Higgs pair (A${}^{0}$h${}^{0}$) shows greater couplings as compared to that of neutral Higgs pair ($A^{0} H^{0}$). Same is case for neutral Higgs triple production ($h^{0} h^{0} A^{0}$) an enhancement is seen in cross-sections with increase of tan{\ss}. So, with the increase of tan{\ss} value the corresponding cross-section value in fb also increases due to greater coupling of neutral triple Higgs ($h^{0} h^{0} A^{0}$) for  neutral Higgs triple production ($h^{0} h^{0} A^{0}$).

\end{document}